\documentclass[a4paper,11pt]{article}
\pdfoutput=1 

\usepackage{jcappub} 
\usepackage[utf8]{inputenc}
\usepackage[T1]{fontenc} 
\usepackage{natbib} 
\usepackage[sort&compress,numbers]{natbib}
\usepackage{multirow}
\usepackage{threeparttable}
\usepackage[]{algorithm2e}
\usepackage{mathtools}

\newcommand{\rel}{\sigma_{\text{rel}}}
\newcommand{\f}{\text{fid}}
\newcommand{\dd}{\mathrm{d}}

\title{A general reconstruction of the recent expansion history of the universe}

\author[a]{S. D. P. Vitenti}
\author[b,c]{M. Penna-Lima}

\affiliation[a]{Institut d'Astrophysique de Paris (${\cal G}\mathbb{R}\varepsilon\mathbb{C}{\cal O}$), \\ UMR7095 CNRS, 98 bis boulevard Arago, 75014 Paris, France}
\affiliation[b]{AstroParticule et Cosmologie (APC), Universit\'e Paris Diderot, \\ UMR7164 CNRS/IN2P3, 10 rue Alice Domon et L\'eonie Duquet, 75013 Paris, France}
\affiliation[c]{Instituto Nacional de Pesquisas Espaciais, Divis\~ao de Astrof\'isica, \\ Av. dos Astronautas 1758, 12227-010, S\~ao Jos\'e dos Campos -- SP, Brazil}

\emailAdd{dias@iap.fr}
\emailAdd{pennal@apc.in2p3.fr}

\abstract{
Distance measurements are currently the most powerful tool to study the expansion history of the universe without specifying its matter content nor any theory of gravitation. 
Assuming only an isotropic, homogeneous and flat universe, in this work we introduce a model-independent method to reconstruct directly the deceleration function via a piecewise function. Including a penalty factor, we are able to vary continuously the complexity of the deceleration function from a linear case to an arbitrary $(n+1)$-knots spline interpolation. We carry out a Monte Carlo (MC) analysis to determine the best penalty factor, evaluating the bias-variance trade-off, given the uncertainties of the SDSS-II and SNLS supernova combined sample (JLA), compilations of baryon acoustic oscillation (BAO) and $H(z)$ data. The bias-variance analysis is done for three fiducial models with different features in the deceleration curve. 
We perform the MC analysis generating mock catalogs and computing their best-fit. For each fiducial model, we test different reconstructions using, in each case, more than $10^4$ catalogs in a total of about $5\times 10^5$.  
This investigation proved to be essential in determining the best reconstruction to study these data. We show that, evaluating a single fiducial model, the conclusions about the bias-variance ratio are misleading. We determine the reconstruction method in which the bias represents at most $10\%$ of the total uncertainty. In all statistical analyses, we fit the coefficients of the deceleration function along with four nuisance parameters of the supernova astrophysical model. For the full sample, we also fit $H_0$ and the sound horizon $r_s(z_d)$ at the drag redshift. The bias-variance trade-off analysis shows that, apart from the deceleration function, all other estimators are unbiased. Finally, we apply the Ensemble Sampler Markov Chain Monte Carlo (ESMCMC) method to explore the posterior of the deceleration function up to redshift $1.3$ (using only JLA) and $2.3$ (JLA+BAO+$H(z)$). We obtain that the standard cosmological model agrees within $3\sigma$ level with the reconstructed results in the whole studied redshift intervals. Since our method is calibrated to minimize the bias, the error bars of the reconstructed functions are a good approximation for the total uncertainty.
}


\arxivnumber{}

\begin{document}
\maketitle
\flushbottom

\section{Introduction}
\label{sec:introduction}

Many indications of the accelerated expansion of the universe come from distance measurements, such as the distance modulus of type Ia supernovae (SNe Ia)~\cite{Riess1998, Perlmutter1999}. In the last two decades, several models have been proposed in order to explain this phenomenon and, in general, they can be classified into dynamic and kinematic models. Assuming the general relativity, the first is described by adding a fluid, Dark Energy (DE), in which several propositions provide different DE equation of state (EoS) (for a review, see \cite{Joyce2015} and references therein). Other common dynamic approach is to modify the geometric setting of the gravitational theory instead of the energy-momentum tensor, such as the high-dimensional models \cite{Dvali2000} and $f(R)$ theories \cite{Sotiriou2010, Nojiri2011}. These approaches are labeled as dynamic in the sense that there are differential equations of motion for the metric, whose modifications consist in altering the source term or the equation of motion itself.

In the context of kinematic models, the expansion history of the universe can be probed without assuming any theory of gravitation nor its matter content, and one only needs to define the space-time metric to study it. Considering the Friedmann-Lema\^{\i}tre-Robertson-Walker (FLRW) metric, the recent expansion of the universe is described in terms of the scale factor and its $n$-order derivatives with respect to time, such as the Hubble, deceleration and jerk functions \cite{Turner2002c, Visser2004, Rapetti2007, Zhai2013}, as well as in terms of the luminosity distance \cite{Daly2003,Benitez-Herrera2012} (and references therein).\footnote{There are also works which explore the properties of the DE equation of state (i.e., assuming the general relativity) using kinematic quantities, e.g., \cite{Huterer2003,Daly2008}.} 

Since the only unknown in this metric is the scale factor (and possibly the spatial curvature), once one of the kinematic functions is determined, the others can be found by integrating and/or differentiating it. Therefore, all kinematic functions are related. Which function will be chosen to be reconstructed depends on the questions one wants to answer.\footnote{For a discussion about the issues in deriving a kinematic function from a reconstructed one of another quantity, see~\cite{Sahni2006}.} For example, one can model the luminosity distance by a linear piecewise function and obtain a statistically sound and unbiased fit using observational data. However, this study will not contribute at all to the understanding of the recent accelerated expansion since, in this case, the deceleration function is assumed zero in the entire redshift interval.\footnote{The second derivative of the distance, given by a linear spline, is zero everywhere.}

Keeping the above idea in mind, the reconstruction of a kinematic function has been addressed using different methods. To understand some of these different methodologies, we divide the problem in two parts: (i) the theory underlying the observable quantities and (ii) the relation between the observables and the data, including the data probability distribution.

Regarding (i), the kinematic model provides a perfect description of the observables, not taking into account any noises nor errors in the measurements. For the sake of argument, suppose that (ii) is not part of the problem, i.e., the data is perfectly known. In this case, we could use a parametric function and adjust its parameters, such that the observable (hopefully) matches the data points, or we could use the data points to determine the observable function using, for example, interpolation. In this sense, we say that the analysis is model-dependent when a given parametric function is chosen a priori and model-independent when we use the data to determine it. 

There are also two main procedures to treat (ii). We can assume which is the probability distribution of the data and, consequently, the only problem left is to determine the observable curve, which can be done in a model-dependent or independent way, as discussed above. In statistics texts, this is described as a parametric method. On the other hand, we can follow an even more conservative path and not impose a given probability distribution for the data. This way, known as non-parametric, also uses the data to reconstruct their own probability distribution. 

In the model-dependent parametric approach, one assumes a priori a specific functional form of kinematic quantities, such as the deceleration function $q(z)$, and a probability distribution of the data~\cite{Riess2004, Shapiro2006, Avgoustidis2009, Nair2012}. A feature of this strategy is that its results have potentially smaller error bars when compared to the others. After all, one is introducing a reasonable set of assumptions which can lead to biased results. A natural improvement to this is to apply a model-independent approach, where one tries to reconstruct the curve when still using the assumed distribution for the data. Among these approaches is the Principal Component Analyses (PCA), in which the kinematic function is described in terms of a set of basis functions and the data is used to determine which subset of this basis is better constrained. Then, the function is reconstructed by using this subset~\cite{Huterer2003, Shapiro2006, Clarkson2010, Ishida2011, Ruiz2012, Benitez-Herrera2013}.

Another possibility is to use smoothing methods~\cite{Shafieloo2006, Shafieloo2012}. In this model-independent non-parametric case, only mild assumptions are made about the data and, usually, no assumption is made about the model. This allows a direct translation of the data into a kinematic curve. Still in this context, we also have the Gaussian Process (GP), in which one chooses to model directly the probability distribution of the kinematic function itself~\cite{Holsclaw2010,Seikel2012}. For a more complete list of non-parametric methods see~\cite{Montiel2014} and references therein.

Recovering both the probability distribution of the data and a reconstructed kinematic function require a large amount of data and, in practice, the current observational cosmology did not seem to have reached this level yet. This is evinced by the results obtained so far in the literature~\cite{Ruiz2012, Shafieloo2012, Seikel2012, Montiel2014}. Regarding the data, there is a good perspective to increasingly improve their probabilistic descriptions, since different error sources, such as the systematic ones, are being included in their modeling (e.g., \cite{Conley2011, Betoule2014} ). This presents an additional challenge to the non-parametric methods, as they must incorporate all the error sources in their reconstruction.

Even in a model-independent and non-parametric approach, the estimated curves are not free from assumptions. Each method has some internal choices of parameters. Currently in the literature, these parameters are obtained using the observational data. However, as we usually have only one set of data, doing so will calibrate the method for this one particular realization of the data. In this case, there is no way to know if this calibration provides the best balance between bias and variance. This difficulty can be circumvented using different realizations of the \textbf{same} data set. For a given calibration, i.e., for a given choice of the internal parameters, the method is applied to a large number of simulations obtaining the bias and variance for this calibration. Then, repeating this process for different calibrations one can find the best suited one for the chosen data set. In other words, the internal parameters in these reconstructions must not be related to one particular realization of the data, but to their probability distribution.

This idea can be extended to the study of the statistical properties of the data. For example, in~\cite{Montiel2014}, among other results, the authors apply a bootstrap-like procedure to calibrate the smoothing parameter applied to the data. This kind of analysis can provide a insightful information about the statistical properties of the data when little is known about their relationships.

In this work, we use the current available observational data for small redshifts ($z \lesssim 2.3$) and their likelihoods, namely, the Sloan Digital Sky Survey-II and Supernova Legacy Survey 3 years (SDSS-II/SNLS3) combined Joint Light-curve Analysis (JLA) SNe Ia sample \cite{Betoule2014}, baryon acoustic oscillation (BAO) data~\cite{Beutler2011, Padmanabhan2012, Kazin2014,  Ross2014} and $H(z)$ measurements~\cite{Stern2010, Riess2011, Moresco2012, Busca2013}. Currently, there is not enough data to perform a full model-independent and non-parametric reconstruction of the recent evolution of the universe. Therefore, we use the usual likelihood for these data, but, to be conservative, we reconstruct $q(z)$ along with some astrophysical parameters of SN Ia, the drag scale (present in the BAO likelihood) and the Hubble parameter $H_0$.

Besides the above data, there is also a wealth of data concerning the large scale structure connected to the perturbations around a FLRW metric, such as the temperature fluctuations of the cosmic microwave background (CMB) \cite{Hinshaw2013, Planck2015}. Since we assume no dynamic model, we would have to propose a kinematic one for the perturbations. Such model is not feasible as it would require a set of functions of both time and space. In principle, one could also directly use derived observables, as, e.g., the CMB distance priors \cite{Komatsu2011}, to fit the background model. However, these parameters are obtained in the context of a specific model, e.g., $\Lambda$CDM. Thus this model would be indirectly reintroduced in the results. For this reason, we choose not to use these data and focus on the distance-like measurements.

We reconstruct $q(z)$ using a model-independent parametric approach, which consists in describing $q(z)$ by a cubic spline. The choice of the deceleration function comes from the fact that one can use such function to directly test the energy conditions~\cite{Lima2008,Lima2008a}. Likewise, the deceleration function is related to the underlying dynamics of the metric, since it is a simple combination of first and second derivatives of the scale factor. Therefore, their knowledge is necessary to constrain models which chooses a different dynamic for the gravitation sector (Alves et~al. to be submitted). In addition, the spline method allows us to vary the complexity of the functional form as a function of the knots number, for example. Here we introduce a novel method to continuously vary the complexity of the reconstructed function. As a result of this analysis, we also obtain the reconstruction of those SN Ia parameters and the drag scale. 

The paper is organized as follows. In section~\ref{sec:def_q} we review basic concepts, such as $q(z)$, and the few assumptions made here. In section~\ref{sec:review}, we discuss some approaches used in the literature and some of their respective drawbacks. In section~\ref{sec:reconst} we present our reconstruction method and the tools to vary the function complexity and to select the best one. Following, section~\ref{sec:data}, we specify the observational data sets used in our study and their respective likelihood functions. We then perform a Monte Carlo analysis in different scenarios calibrating the method (section~\ref{sec:meth_valid}). Finally, we use this calibration to reconstruct the deceleration function using the observational data, in section~\ref{sec:real_reconst}, and we summarize our conclusions in section~\ref{sec:conclusions}.

\section{Deceleration function $q(z)$}
\label{sec:def_q}

Measurements of the CMB \cite{Hinshaw2013, Planck2015} show that the universe is nearly homogeneous and isotropic at large scales. Therefore, we assume that the universe follows the cosmological principle, restricting the metric to the FLRW metric,
\begin{equation}
\label{eq:rw_metric} 
ds^2 = -c^2\,dt^2 + a^2 (t) \left [ dr^2 + S_k^2(r) (d\theta^2 + \sin^2 \theta  d\phi^2) \right ]\;,
\end{equation}
where $S_k(r)=(r\,$, $\sin(r)$, $\sinh(r))$ for flat, spherical and hyperbolic spatial section ($k=0, 1, -1$), respectively, $c$ is the speed of light and $a(t)$ is the cosmological scale factor. In this case, the expansion history of the universe can be defined knowing $a(t)$ and $k$.

In practice, we do not measure $a(t)$ directly, but related quantities such as the distances to astronomical objects. Considering a null trajectory of photons emitted by a galaxy traveling along the radial direction to us, we have that
\begin{equation}\label{eq:com_dist1}
r = c \int_{t_e}^{t_0} \frac{\dd t^\prime}{a(t^\prime)},
\end{equation}
where $t_e$ and $t_0$ are the emitted and observed times, respectively. 
Expanding the scale factor to second order around $t_0$, gives
\begin{equation}
a(t) = a_0 + H_0 (t - t_0) - \frac{q_0 H_0^2}{2} (t - t_0)^2,
\end{equation} 
where $a_0$ is the scale factor today and $H_0$ and $q_0$ are, respectively, the Hubble and deceleration functions at $t_0$, 
\begin{equation}
H_0 = \left(\frac{ \dot{a}}{a} \right)_{t_0} \quad \text{and} \quad q_0 \equiv - \left(\frac{\ddot{a}a}{\dot{a}^2}\right)_{t_0}.
\end{equation}
Rewriting the comoving distance $D_c$ and the deceleration function in terms of the redshift, $1 + z = a_0/a$, we obtain
\begin{equation}\label{eq:com_dist}
D_c(z) = a_0 r = \frac{c}{H_0} \int_0^z \frac{\dd z^\prime}{E(z^\prime)}
\end{equation}
and
\begin{equation}\label{eq:qz}
q(z) = \frac{(1 + z)}{H(z)}\frac{\mathrm{d}H(z)}{\mathrm{d}z} - 1,
\end{equation}
whose integral solution is
\begin{equation}\label{eq:Ez}
E(z) = \exp{\int_0^z \frac{1+q(z^\prime)}{1+z^\prime}\dd z^\prime},
\end{equation} 
and where $E(z)$ is the normalized Hubble function, $E(z) = H(z)/H_0$. 

Equations~\eqref{eq:com_dist} -- \eqref{eq:Ez} evince that we shall reconstruct $q(z)$ in order to access local information about the accelerated/decelerated phase~\cite{Lima2008}. Besides, assuming that $q(z)$ is continuous guarantees that $E(z)$ and $D_c(z)$ are also continuous functions and at least once and twice differentiable, respectively, although the opposite is not true.

\section{Review of other approaches}
\label{sec:review}

In this section we briefly present some of the most used methods in reconstructing the expansion history of the universe. We discuss some intrinsic issues of these methodologies, which motivated us to develop a novel approach (presented in section~\ref{sec:reconst}). 

\subsection{Parametric models}
\label{sec:PM}

In general the Taylor series approaches have two related problems. The first is the convergence radius of the Taylor expansion itself, which can only be estimated since the real scale factor is naturally unknown. On the other hand, as we are fitting the coefficients of such expansion, the functional form for the scale factor (or distance) can be interpreted as a simple polynomial interpolation. In this sense, changing the time parameter can be useful \cite{Cattoeen2008} and provide a better polynomial interpolation.
However, this leads us to the second problem, the Runge's phenomenon. That is, after a given order, higher order polynomials provide worse and worse approximations. Therefore, when using the Taylor expansion one should stay on a small convergence radius, which would restrain the analysis to a very small but unknown redshift or use a polynomial interpolation keeping in mind its caveats. 

More generally, the problem of finding a good kinematic description of the expansion history can be addressed using a parametric method. In this context, one assumes a priori a specific functional form of a kinematic function, like the polynomial form discussed above, then proceeds by fitting its parameters using observational data. For example, in Refs.~\cite{Riess2004, Shapiro2006, Avgoustidis2009, Nair2012} they fit different functional forms of the deceleration function $q(z)$. The drawback of this method is that the choice of a functional form introduces a form-bias in the estimates if the functional form is different from the true one. As we do not know it, the result of such fit can be misleading since, even if the parameters' error bars are small, their form-biases can still be large. 
A more conservative approach is to use flexible functional forms. However, this translates in using many parameters and, consequently, obtaining larger error bars. In this way, there is a natural trade-off between variance and form-bias which should be evaluated to determine the optimal reconstruction.

An additional, less discussed, difficulty is the estimator-bias. The functional forms are usually fitted using a Least-Squares (LS) or Maximum-Likelihood (ML) approach. As it is well known, both approaches can provide biased estimators for the parameters (ML estimators are usually asymptotically unbiased). This means that, even if we knew the correct functional form, the fact that we have only a finite number of observations can lead to estimator-biases.\footnote{For a detailed discussion of this problem in the context of the cluster number counts see~\cite{Penna-Lima2014}.} Heuristically, when fitting a functional form with $n$ parameters using $m$ data points we will have $m/n$ observations per point. Hence, if the estimators are only asymptotically unbiased, then the higher the number of parameters higher the estimator-biases. Thus, the final trade-off must consider variance, form-bias and estimator-bias.

\subsection{Principal Component Analysis} 
\label{sec:PCA}

Another popular methodology is the Principal Component Analysis (PCA) \cite{Huterer2003, Shapiro2006, Clarkson2010, Ishida2011, Ruiz2012}. In this approach the kinematic function is described by a flexible parametrization (e.g., in~\cite{Huterer2003} they express the DE EoS as a constant piecewise function within $N=50$ bins). From the estimated covariance matrix of these parameters, the eigenvectors and eigenvalues are deduced. Say, for example, that we choose the first 5 eigenvectors, whose eigenvalues correspond to the smallest variance terms. This means that the original division in 50 bins is being described by a 5-dimensional parametrization.

The eigenvectors, whose eigenvalues correspond to the smallest variance terms, provide the parametrization which is better constrained by the data. The appeal of this method is that it provides a straightforward way to determine the curves better constrained by the data. The choice of how many eigenvectors should be used can be answered by looking the variance-bias trade-off \cite{Huterer2003}.
On the other hand, the method as described above is subject to a potential drawback. The relationship between different kinematic functions are given in terms of their integral in time, for example, the relation between the deceleration and the Hubble functions is given by Eq.~\eqref{eq:Ez} while the distance is another integral of the Hubble function [Eq.~\eqref{eq:com_dist}]. Note that, to calculate the distance to a given SNIa at redshift $z_i$, the deceleration function is integrated twice between $[0, z_i]$. Therefore, all bins in this interval contribute to the final value of the distance. This leads to the following problem, the first bins contribute to the value of the distance for almost all supernovae, while the final bins affect only few objects. Besides, since the initial bins always contribute to the value of the distance at higher redshifts, they will be naturally correlated to them, any modification in their value will have to be compensated by the other. These facts have the following consequence, the best constrained modes will be strongly connected to the first bins while the worst will be related mostly to the last bins. Therefore, when one chooses to use only a few (better constrained) eigenvectors, the final parametrization will provide almost no power in the last bins. This is natural since it is equivalent to choose the coefficients of the last eigenvectors to be fixed at zero. This problem can be seen in~\cite{Huterer2003}, where the issue persists even when the authors consider a forecast with 3000 SNe Ia uniformly distributed in redshift. This problem was also noted in~\cite{Ishida2011}. In the latter, they realized that the first modes have this deficiency and proposed the addition of a new parameter to circumvent this problem. 

Similarly to what is commonly used in the PCA approach, the authors in~\cite{Crittenden2012,Zhao2012} reconstructed the EoS function describing it as a set of bins (discontinuous piecewise constant function of the scale factor). Instead of limiting the estimated curve variance by using a small set of eigenvectors (principal components), they introduced the correlated prior, in an approach similar to the GP, which treats the reconstructed curve as a set of random variables emerging from a multidimensional Gaussian distribution. This prior introduces a correlation between the bins controlled by the correlation length and prior strength. Once the prior is calibrated, the problem described in the above paragraph can be systematically resolved. 

The discussion above elucidates an important characteristic of the PCA, it is dependent on the initial description of the function. In many works the simple step function (or bins) were used. If another basis functions were used, with different behavior at large redshifts, we would end up with a similar problem, i.e., the large redshift behavior of the best constrained basis function would dominate at large redshifts, providing biased estimates at these points. We conclude that, in the specific context of constraining a function by its integral, the PCA approach has this potential problem.

\subsection{Smoothing methods}
\label{sec:SM}

As we mentioned in section~\ref{sec:introduction}, these approaches are model-independent and non-parametric. 
Therefore, they suit the study cases where the data probability distribution is unknown. However, as pointed out by Montiel et~al.~\cite{Montiel2014}, these methods find difficulties due to the limited amount of observational data or even due to some features of the methods themselves, such as the size of the smoothing parameter and the assumptions on priors or fiducial models. Besides, since no assumption is made about the data distribution, one cannot use resampling to perform a self-validation, but only bootstrap like procedures, e.g., jackknife, which usually requires large samples. 

Finally, the fact that this method makes such minimal assumptions is not necessarily useful. As one can only reconstruct the direct variable associated to the observable, i.e., cosmological distances and $H(z)$, any inference about the derived kinematic quantities is limited since it is highly dependent on the smoothing technique as, e.g., smoothing linear splines has always zero second derivative, and top-hat moving average filters are non-continuous. Therefore, any analysis about kinematic quantities, different of the reconstructed one, requires new assumptions~\cite{Daly2004, Shafieloo2012}.

\subsection{Gaussian Process}
\label{sec:GP}

In this approach, instead of modeling a kinematic function, one chooses to model a probability distribution for the curve as a Gaussian probability distribution. This assumption dictates the data probability distribution by relating both the curve and observable probability distributions. In this sense, this approach unifies the two aspects of the model, the data distribution and the curve reconstruction. All the assumptions are comprised in the mean curve and the two-point covariance, which define the Gaussian distribution of the curve. 

The drawback is similar to the parametric procedure. In general, one has to assume a function to describe the mean of the GP and a two-point function to describe the variance. If the considered mean function differs from the true one, it will impose a bias in the reconstruction. See, for example, references~\cite{Holsclaw2010, Seikel2012}, where they assume a constant mean function. Since the GP determines the observable statistical distribution, it is also necessary to include the data distribution to calculate the joint probability distribution of both curve and data. In this case, one has a model-independent but parametric method  in the sense that one is assuming a particular distribution for the data. The GP validation also has to be performed for a number of realizations of the data, since the indirect curve determination through a Gaussian distribution can lead to bias in an unpredictable way.

\section{Reconstruction of $q(z)$}
\label{sec:reconst}

Broadly speaking each strategy described in section~\ref{sec:review} has a better suited application. As a rule of thumb, to use less hypothesis it is necessary to have more data. Therefore, if the amount of data is limited, the generality of GP and soothing methods, for example, are restricted, leading to underdetermined problems. Besides that, including natural hypothesis can also be considerably difficult in those non-parametric studies, e.g., after the determination of the cosmological distances through a smoothing method it is necessary to add new assumptions to describe the Hubble function.

Therefore, in this work we adopted a model-independent parametric approach. Nonetheless, to be conservative we reconstruct the kinematic curve along with all the phenomenological parameters related to the modeling of each data set. Doing so, we minimize the assumptions on the data distribution bypassing any bias which could result from it. In a model-independent technique we need to use a set of functions to perform the reconstruction. To avoid the problems described above on the PCA approach, we use a cubic spline to reconstruct our observable, as described in section~\ref{sec:spline}. One advantage of the cubic spline is that it is continuous and twice differentiable on every knot, thus, all the parameters (the value of the function on each knot) are related through these conditions.\footnote{In practice, once the value of the function at each knot is defined, the coefficients of each cubic polynomial are obtained from the solution of a linear system including not only the knots on its neighborhood but all knots.}

Another point on the model-independent approach is how to deal with the bias. In the PCA approach one uses the particular set of data to determine which basis functions form the minimal set. Depending on the data distribution, this tends to favor the intervals where the sample is denser, which usually creates biases on the other regions. To avoid this issue, we impose a global penalization (see section~\ref{sec:penalty}) on the curvature of the curve. Applying the same penalization in each interval, we evade the localization problem described before.

\subsection{Piecewise deceleration function}
\label{sec:spline}

In this work, we avoid making arbitrary choices of the $q(z)$ form, and, consequently, \textit{a priori} restricting it to specific functional forms, by approximating $q(z)$ by a piecewise third-order polynomial function, i.e., a cubic spline. 

The first step to build an estimator of $q(z)$ [denoted as $\hat{q}(z)$] is to specify the redshift interval (domain $\mathrm{D}$) in which the function is defined. This interval is $\mathrm{D} = [z_{min}, z_{max}]$, where $z_{min}$ ans $z_{max}$ are the minimum and maximum redshifts of the used data. The next step is to choose the partition of the domain $\mathrm{D}$ into $n$ sub-intervals in which we define $\hat{q}(z)$ as a cubic polynomial function, namely,
\[\hat{q}(z) = \left\{ \begin{array} {llllll}
             p_0(z) = a_0(z-z_0)^3 + b_0(z-z_0)^2 + c_0(z-z_0) + d_0 \quad & {z \in [z_0,z_1)}\\
             p_1(z) = a_1(z-z_1)^3 + b_1(z-z_1)^2 + c_1(z-z_1) + d_1 & {z \in [z_1,z_2)}\\
             \qquad\qquad\qquad\qquad \vdots & \quad \vdots\\       
             p_{n-1}(z) = a_{n-1}(z-z_{n-1})^3 + b_{n-1}(z-z_{n-1})^2 + c_{n-1}(z-z_{n-1}) + d_{n-1} & z \in [z_{n-1}, z_n],
     \end{array} \right.
\]
where $z_0 = z_{min}$, $z_n = z_{max}$ and $p_i$ is the cubic polynomial defined in the $i$-th sub-interval. Note that each polynomial $p_i(z)$ in the segment $[z_i,z_{i+1})$ depends on 4 parameters ($a_i$, $b_i$, $c_i$ and $d_i$) and, consequently, we would need to estimate $4n$ parameters to define $\hat{q}(z)$ in the whole domain $\mathrm{D}$. However, imposing the following continuity conditions, 
\begin{eqnarray*}
p_{i+1}(z_{i+1}) &=& p_i(z_{i+1}),  \\
p^{\prime}_{i+1}(z_{i+1}) &=& p^{\prime}_i(z_{i+1}), \\
\quad p^{\prime\prime}_{i+1}(z_{i+1}) &=& p^{\prime\prime}_i(z_{i+1}),
\end{eqnarray*}
on the $n-1$ internal knots $i \in (1,n-1)$, where ${}^\prime$ denotes the derivative with respect to $z$, and the two not-a-knot boundary conditions 
\begin{equation}
\hat{q}^{\prime\prime\prime}_{0}(z_{1}) = \hat{q}^{\prime\prime\prime}_1(z_{1}) \quad \text{and} \quad \hat{q}^{\prime\prime\prime}_{n-2}(z_{n-1}) = \hat{q}^{\prime\prime\prime}_{n-1}(z_{n-1}),
\end{equation}
we end up with only $n + 1$ parameters to determine $\vec{Q} = \{d_0, ..., d_{n}\}$, i.e., the values of $\hat{q}(z)$ at each knot. As we have a one to one relation between the function in each knot $\hat{q}_i \equiv \hat{q}(z_i)$ and the parameter $d_i$ of each polynomial, from now on we rename the set as $\vec{Q} = \{\hat{q}_i\}$ for the sake of notational simplicity. Thus, using this cubic spline approximation, we fit the vector $\vec{Q}$ in order to obtain the estimates of the deceleration function (see description in section~\ref{sec:meth_valid}).  

Any interpolation method introduces an error source limiting the set of functions able to be reconstructed. The interpolation error for a cubic spline has an upper bound proportional to both the largest distance between nearby knots to the fourth power and the fourth derivative of the function with respect to $z$ (for details see~\cite{Boor2001}). At first sight, larger the number of knots, smaller the interpolation error. But in practice, the number of knots is limited by the increasing number of parameters. Besides, there is also the overfitting that rises when fitting the parameters using data points. As a rule of thumb, one should choose $n$ and, consequently, the intervals between knots, such that the estimated function is expected to be well approximated by a cubic polynomial in these intervals. One can test the choice of $n = n_1$ applying the reconstruction for another one, e.g., $n = n_1 + 1$, and probing the results for any significant improvements on the fit. Notwithstanding the interpolation error, another important source of uncertainty comes from the statistical errors (bias/overfitting), as we will discuss in the next section. Finally, the use of cubic splines represents a large advantage in comparison to the step functions frequently used in the literature. For a constant piecewise function, the interpolation error is bounded by the first derivative times the largest distance between nearby knots. As a result, the number of knots (and, consequently, parameters) necessary to reconstruct functions with the same interpolation error bound is much larger in a binned approach.


\subsection{Function complexity}
\label{sec:penalty}

Assuming a cubic spline to approximate $q(z)$, we are able to address both model-dependent and model-independent parametric methods by varying the number of knots. The simplest case, $n = 4$, is equivalent to consider that $q(z)$ is a third-order polynomial. On the other hand, we approach a model-independent case increasing $n$. The complexity of the function $\hat{q}(z)$ is, in principle, parameterized by the number of knots, as the number of knots goes to infinity any interpolation error drops to zero. Nonetheless, the choice of the domain partition is rather arbitrary and, at first, one would have to test different options in order to achieve, for example, a ``model-independent limit'' trying to minimize the over-fitting error. Another difficulty inherent of this approach is that the number of knots is a discrete variable and, as such, it is difficult to include it as another parameter in the analysis. 

Instead of varying the number of knots by adding/removing actual knots to the function representation, we can fix the number of knots in some large value and penalize independent values of the parameters. For example, given that the parameters are just the values of the function at each knot, a penalty factor as a increasing function of $\vert\hat{q}_i - \hat{q}_{i+1}\vert$ will correlate all parameters $\hat{q}_i$. The correlation will be proportional to the weight of the penalization in the analysis, higher the weight more correlated are the parameters. In the strong correlation limit, all parameters would be equal, i.e., $\hat{q}_i = \hat{q}_{i+1}$. In this last case, even with a large number of knots, the effective number of degrees of freedom would be one. In short, varying the weight of the penalization, we can vary the effective number of degrees of freedom, circumventing the difficulties described above.\footnote{One can also easily interpret, using the Bayesian point of view, the penalty factor as a prior on the fitted function.}

\begin{figure}
\begin{center}
\includegraphics[scale=0.7]{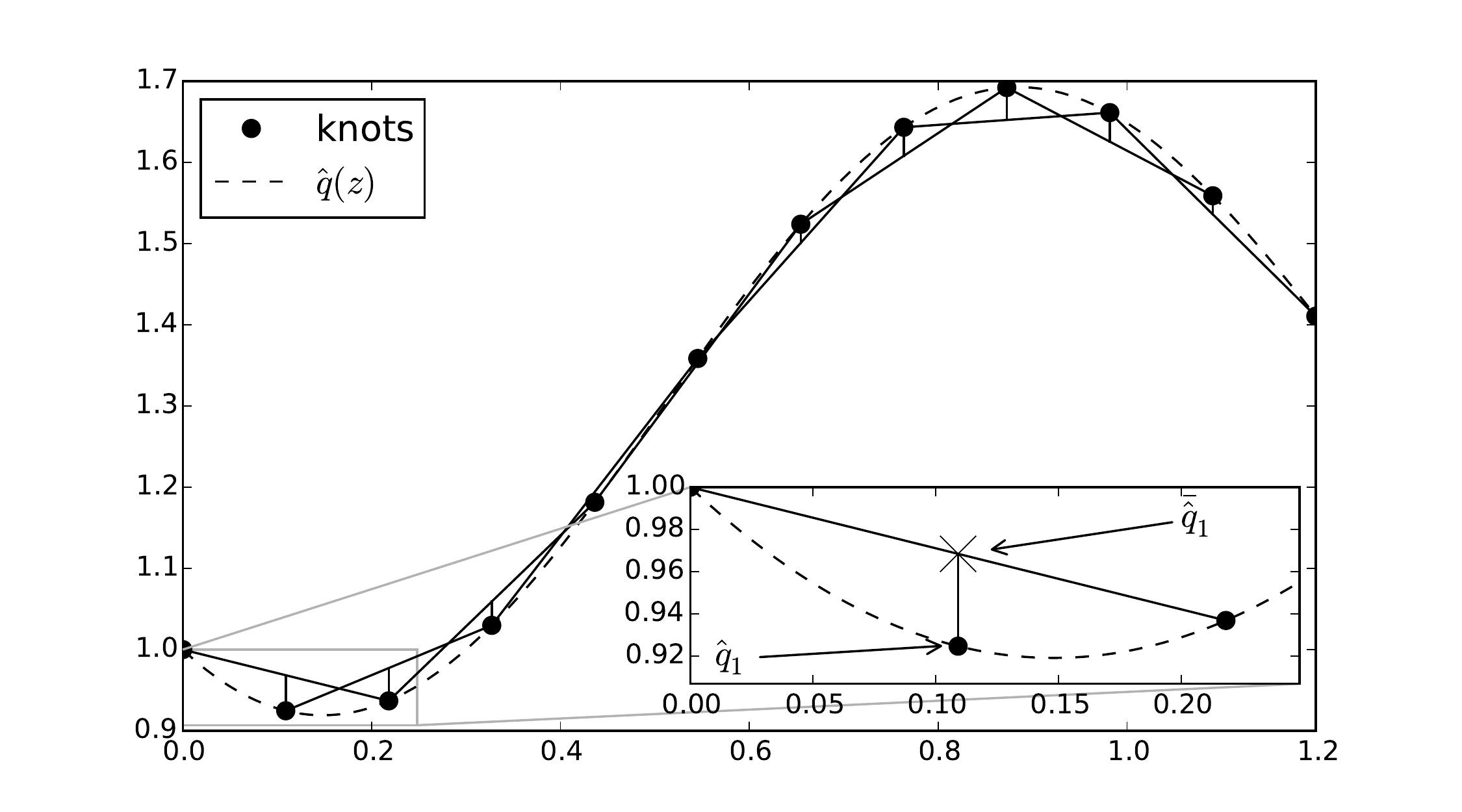}
\end{center}
\caption{\label{fig:basis} A descriptive example of the penalization on a cubic spline. For each three knots, the penalization is proportional to the distance between the straight line connecting the first and third knots, and the interpolation function.}
\end{figure}

In practice, we include a set of penalty factors $P_i(\sigma_i)$ in our estimator. The initial likelihood is $L(\vec{D}, \vec{\theta})$, where the vector $\vec{D}$ represents the data set and the vector $\vec{\theta}$ all the parameters, including the spline parameters $\hat{q}_i$ and other parameters as described in sections~\ref{sec:mc_sneia} and \ref{sec:mc_alldata}. To obtain the parameter estimators, we add to the likelihood $L(\vec{D}, \vec{\theta})$ the penalization $P_i(\sigma_i)$ defining the penalized likelihood
\begin{equation}\label{eq:likel_penalty}
-2 \ln\left[L_P (\vec{D}, \vec{\theta})\right] \equiv -2 \ln\left[L\left(\vec{D}, \vec{\theta}\right)\right] + \sum_{i = 2}^{n-1} P_i(\sigma_i),
\end{equation}
where the penalty factor is given by 
\begin{equation}
\label{eq:penalty_fac}
P_i(\sigma_i) = \left(\frac{{\bar{\hat{q}}_i - \hat{q}_i}}{\sigma_i}\right)^2,\qquad \bar{\hat{q}}_i = \frac{(\hat{q}_{i-1} + \hat{q}_{i+1})}{2}, \qquad \sigma_i = \sigma_{abs} + \bar{\hat{q}}_i\rel,
\end{equation}
and we use $\sigma_{abs} = 10^{-5}$.\footnote{The $\sigma_{abs}$ factor guarantees that, even if some $\bar{\hat{q}}_i \simeq 0$ the denominator in the penalty factor does not go to zero.} The penalization factor is schematically illustrated in figure~\ref{fig:basis} showing the positions of $\hat{q}_1$ and $\bar{\hat{q}}_1$. 

We control the complexity of $\hat{q}(z)$ by varying the value of the relative error $\rel$. For example, we are able to recover a high complexity function, in particular, a full $n+1$ knots spline for large $\rel$, and a straight line in the entire redshift interval when $\rel$ goes to zero. The former has many coefficients and can tend to fit the data noise, i.e., it is over-fitting dominated. The second naturally sharpen the constraints on $\{\hat{q}_i\}$, but they can be biased if the assumed functional form significantly differs from the true one. It is worth mentioning that this penalty factor allows us to explore a wide range of functional forms, since its simplest case is a linear function. Meanwhile, without using the penalty factor, the simplest model would be a third-order polynomial.

Finally, we emphasize that, in principle, one could use a large set of knots while constraining the allowed shape with the penalty factor. The restriction will be practical, the computational cost increases with the number of knots. Therefore, one should find the best balance between computational cost and flexibility of the method. 

\subsection{Bias-variance trade-off}
\label{sec:bv_trade}

Giving the penalty function which allows us to explore different complexity forms of $\hat{q}(z)$, we now have to introduce a criterion to determine the best $\rel$ value. We want to find the scenario where the combined error due to the biases (estimator- and form-bias) and the over-fitting error is minimized. For this, we decompose the error into bias and variance components as described below~\cite{Bishop1996, Keijzer2000, Wasserman2001}.

We create a controlled environment introducing a fidicual deceleration function $q^{\f}(z)$, which is determined by a given set of values for the parameters $\vec{\theta}^{\f}$. The idea is to use this function to generate a new data set $\vec{D}$, which is possible since we know the likelihood of the data $L(\vec{D},\vec{\theta})$. We define the ML estimators using the penalized likelihood, then, given a simulated sample $\vec{D}^{(l)}$, we obtain the estimates $\tilde{\vec{\theta}}(\vec{D}^{(l)})$ computing $$\frac{\partial L_P}{\partial \vec{\theta}}\left(\vec{D}^{(l)}, \tilde{\vec{\theta}}\right) = 0.$$ This provides an implicit definition of the function $\vec{\theta}^{(l)} \equiv \tilde{\vec{\theta}}(\vec{D}^{(l)})$. In principle, we could calculate the bias in the estimator integrating the function $\vec{\theta}^{(l)}$, i.e., $$\left\langle\tilde{\vec{\theta}}\right\rangle = \int\dd\vec{D} \, \vec{\theta}^{(l)}L_P\left(\vec{D},\vec{\theta}^{\f}\right).$$ However, such integration is computationally unfeasible. It is a $N$ dimensional integration, where $N$ is the number of data points, and the function $\vec{\theta}^{(l)}$ is usually only determined numerically by maximizing the penalized likelihood.

Instead, we use the Monte Carlo (MC) approach to deal with such integrals.\footnote{Note that here we are using the MC method to perform integrals in the data $\vec{D}$ given a set of parameters $\vec{\theta}^\f$. Therefore, we are sampling new data from the given likelihood. This procedure is similar but has different applications than the MC used to study the parameter space $\vec{\theta}$ given a data set $\vec{D}$.} Since we know the probability distribution of the data, we can create a new realization of the data (mock catalog) $\vec{D}^{(l)}$ by resampling,\footnote{The likelihoods used in this work were all implemented on top of the data description objects of the Numerical Cosmology library~\cite{DiasPintoVitenti2014}. These objects automatically provide the resample feature for any likelihood using them.} i.e., using a (pseudo)random number generator to create a new data set. Given a large enough number of realizations $m$, we can approximate the expected value of a function of the data as
\begin{equation}
\left\langle f\left(\vec{D}\right) \right\rangle \simeq \frac{1}{m} \sum_{l=1}^{m} f\left(\vec{D}^{(l)}\right).
\end{equation}

Using these tools, we introduce the mean squared error (MSE) of approximating a fiducial function $q^{\f}(z)$ by $\hat{q}(z; \rel)$. For a fixed $\rel$, it is
\begin{equation}
\text{MSE} = \left\langle\left[\hat{q}^{(l)}(z; \rel) - q^{\f}(z)\right]^2\right\rangle \simeq \frac{1}{m} \sum_{l=1}^{m} \left[\hat{q}^{(l)}(z; \rel) - q^{\f}(z)\right]^2.
\end{equation}
Given the estimate of the expected value,
\begin{equation}\label{eq:expec_val}
\langle \hat{q}(z, \rel) \rangle \simeq \frac{1}{m} \sum_{l=1}^{m} \hat{q}^{(l)}(z; \rel),
\end{equation}
we have that (omitting $\rel$ for simplicity)
\begin{align}
\left[\hat{q}^{(l)}(z; \rel) -  q^{\f}(z)\right]^2 &= \left[\hat{q}^{(l)}(z) - \langle \hat{q}(z) \rangle + \langle \hat{q}(z) \rangle -  q^{\f}(z)\right]^2 \nonumber \\
&= \left[\hat{q}^{(l)}(z) - \langle \hat{q}(z) \rangle \right]^2 + \left[\langle \hat{q}(z) \rangle -  q^{\f}(z)\right]^2 \nonumber \\
& + 2 \left[\hat{q}^{(l)}(z) - \langle \hat{q}(z) \rangle\right]\left[\langle \hat{q}(z) \rangle -  q^{\f}(z)\right],
\end{align}
and, therefore,
\begin{equation}\label{eq:mse_1}
\text{MSE} \simeq \left(\langle \hat{q}(z) \rangle -  q^{\f}(z)\right)^2 + \frac{1}{m} \sum_{l=1}^{m} \left(\hat{q}^{(l)}(z) - \langle \hat{q}(z) \rangle \right)^2,
\end{equation}
since
\begin{equation}
\frac{1}{m} \sum_{l=1}^{m} \left(\hat{q}^{(l)}(z) - \langle \hat{q}(z) \rangle\right)\left(\langle \hat{q}(z) \rangle -  q^{\f}(z)\right) \simeq 0.
\end{equation}
The first term of eq.~\eqref{eq:mse_1} is the squared bias $b_{\hat{q}}(z)^2$ and the second is the variance. Since this variance estimator is biased, in this work, we evaluate the bias-variance trade-off computing
\begin{align}\label{eq:mse_2}
\text{MSE} &\simeq \left(\langle \hat{q}(z) \rangle -  q^{\f}(z)\right)^2 + \frac{1}{m -1} \sum_{l=1}^{m} \left(\hat{q}^{(l)}(z) - \langle \hat{q}(z) \rangle\right)^2 \nonumber \\
&\equiv b_{\hat{q}}(z; \rel)^2 + \text{Var}\left(\hat{q}(z; \rel)\right).
\end{align}
Both the squared bias and the variance have the same weight in the above expression. Nonetheless, when applying the reconstruction for real data, we usually do not have access to an estimate of the bias. Therefore, minimizing MSE can potentially lead us to a methodology with a large bias (as we will see in sections~\ref{sec:mc_sneia} and \ref{sec:mc_alldata}). To avoid this, in what follows we will minimize the MSE satisfying the constraint 
\begin{equation}\label{eq:m_b}
\frac{b_{\hat{q}}(z; \rel)}{\sigma(\hat{q}(z; \rel))} \leq m_b, \qquad \sigma(\hat{q}(z; \rel)) \equiv \sqrt{\text{Var}(\hat{q}(z; \rel))},
\end{equation}
where $m_b$ controls the maximum ratio between bias and variance.

The variance of the reconstructed curve $\hat{q}(z; \rel)$ can be written in terms of the covariance of the  spline parameters $\text{Cov}\left(\hat{q}_i, \hat{q}_j\right)$. In turn this covariance can be estimated using the unbiased covariance estimator
\begin{equation}\label{eq:cov_q}
\text{Cov}\left(\hat{q}_i, \hat{q}_j\right) = \frac{1}{m - 1} \sum_{l = 1}^{m} \left(\hat{q}_{i}^{(l)} - \langle \hat{q}_i \rangle\right) \left(\hat{q}_{j}^{(l)} - \langle \hat{q}_j\rangle \right),
\end{equation}
where $\hat{q}_i^{(l)}$ is the best-fitting value of the $i$-th spline parameter using the $l$-th mock catalog.

\section{Observational data}
\label{sec:data}

As $q(z)$ is not a direct observable, we need to use other quantities to access $\{\hat{q}_i\}$. In this section, we present the samples of type Ia SNe, BAO and $H(z)$ measurements, and also their respective likelihood functions that we utilize to recover the deceleration function $q(z)$.

\subsection{Type Ia supernova data}

We use the JLA sample \cite{Betoule2014} of 740 SNe Ia, whose likelihood is
\begin{equation}\label{eq:lik_snia}
-2\ln(L_{SNIa}) = \Delta{}\vec{m}^T\mathsf{C}_{SNIa}^{-1}\Delta{}\vec{m},
\end{equation}
where the data covariance is a combination of the systematic and statistical errors $\mathsf{C}_{SNIa} = \mathsf{C}_{sys} + \mathsf{C}_{stat}(\alpha,\beta)$,  and 
\begin{eqnarray}\label{eq:snia_mb}
\Delta{}m_i &=&  m_{Bi} - m_{Bi}^{\text{th}}\\ 
&=& m_{Bi} -5\log_{10}(\mathcal{D}_L(z^{\text{hel}}_i, z^{{\text{cmb}}}_i)) + \alpha X_i - \beta \mathcal{C}_i - M_{h_i} + 5\log_{10}(c/H_0) - 25. \nonumber
\end{eqnarray}
$m_{Bi}$ is the rest-frame peak B-band magnitude of the $i$-th SN Ia, and $z^{\text{hel}}_i$ and $z^{\text{cmb}}_i$ are its heliocentric and CMB frame redshits, respectively. The SN Ia astrophysical model contains four parameters $(\alpha, \beta, M_1, M_2)$, where the first two are related to the stretch-luminosity and colour-luminosity, respectively, and $M_1$ and $M_2$ are absolute magnitudes.  The luminosity distance~\cite{Davis2011} is 
\begin{eqnarray}\label{eq:dis_SN}
D_L(z^{\text{hel}}, z^{\text{cmb}}) &= \frac{c}{H_0} \mathcal{D}_L(z^{\text{hel}}, z^{\text{cmb}}) \nonumber \\ 
&= (1 + z^{\text{hel}}) D_M(z^{\text{cmb}}),
\end{eqnarray}
where the transverse comoving distance in a flat spatial sections universe is $D_M(z) = D_c(z)$.

\subsection{Baryon acoustic oscillation}

The peak position of the angular correlation function of the matter density can be measured by the distance ratio $D_V (z) / r_s(z_d)$ (see \cite{Thepsuriya2014, Aubourg2014} and references therein).  The volume-averaged-distance for perturbations along and orthogonal to the line of sight is defined as~\cite{Eisenstein2005}
\begin{equation}\label{eq:bao_dv}
D_V(z) \equiv \left[ D_M(z)^2 \frac{cz}{H(z)}\right]^{1/3}.
\end{equation}
The sound horizon $r_s(z_d)$ at the drag redshift $z_d$ (i.e., epoch at which baryons were released from photons) is 
\begin{equation}\label{eq:bao_rd}
r_d \equiv r_s(z_d) = \frac{1}{H_0}\int_{z_d}^\infty dz \frac{c_s(z)}{E(z)},
\end{equation}
where $c_s(z)$ is the sound wave speed in the photon-baryon fluid. 

In this work we use 6 BAO data points as described in table~\ref{tab:bao} in appendix~\ref{app:sampling}. The first is measured by Beutler et~al.~\cite{Beutler2011} using galaxies from the 6dF Galaxy Survey. Padmanabhan et~al.~\cite{Padmanabhan2012} reported an improved data obtained with the reconstruction method~\cite{Eisenstein2007} using Luminous Red Galaxy sample from SDSS Data Release 7 (DR7). Kazin et~al.~\cite{Kazin2014} give three points computing the power spectrum and correlation function of galaxies from the WiggleZ Survey in three correlated redshift bins. The last data is obtained by Ross et al.~\cite{Ross2014} which used galaxies from SDSS DR7 with $z < 0.2$. 

The BAO likelihood is
\begin{equation}\label{eq:lik_bao}
-2 \ln L_{\text{BAO}} = \left(\vec{b}^{\text{th}} - \vec{b} \right)^T\mathsf{C}_{\text{BAO}}^{-1}\left( \vec{b}^{\text{th}} -\vec{b} \right) - 2 \ln L_{\text{Ross}},
\end{equation}
where $\vec{b}^{\text{th}}$ is the observable vector calculated using the theoretical model, i.e., the components are given by $b^{\text{th}}_i = D_V(z_i)/r_d$ calculated at each redshift (second column of table~\ref{tab:bao}). The vector $\vec{b}$ represent the observed version of these quantities and its components are provided by the third column of table~\ref{tab:bao}. The matrix $\mathsf{C}^{-1}_{\text{BAO}}$ is the inverse covariance matrix appearing in the BAO likelihood (see table~\ref{tab:bao}). Finally, Ross et al.~\cite{Ross2014} pointed out that their data should be used considering their estimate of the likelihood distribution [which we called $L_{\text{Ross}}$ in eq.~\eqref{eq:lik_bao}], since, in this case, the Gaussian distribution is not a good approximation. 

References~\cite{Beutler2011} and \cite{Padmanabhan2012} used the Eisenstein \& Hu~\cite{Eisenstein1998} (EH98) fitting function to compute $r_d^{\f}$ while \cite{Kazin2014, Ross2014} used CAMB~\cite{Lewis2000}. As mentioned in Ref.~\cite{Kazin2014}, the difference between $r^{\f}_{d,\text{EH98}}$ and $r^{\f}_{d, \text{CAMB}}$ is of order of $3\%$. So due to the current error magnitude of these data, this difference is relevant and, hence, we have to re-scale the data such that all measurements refer to the same method. In particular, we multiply Beutler and Padamanabhan's data by $r_{d, \text{EH98}}^{\f} / r_{d, \text{CAMB}}^{\f} = 1.027$ and 1.025, respectively.

The BAO observable depends on the kinematic model through $D_V(z)$ and  also requires the $r_d$ value. However, to calculate $r_d$ theoretically, we would need to extend the kinematic model to high redshifts and to compute the decoupling redshift $z_d$. We avoid this making $r_d$ a free parameter in the analysis. Therefore, throughout this work we fit $r_d$ along with the other parameters.

\subsection{Hubble function}

We work with 21 measurements of $H(z)$: 11 are provided by Stern et al.~\cite{Stern2010} in the redshift range $0.1 \leq z \leq 1.75$ obtained from the spectra of red-enveloped galaxies; Riess et al.~\cite{Riess2011} estimated the Hubble constant $H_0$ using optical and infrared observations of over 600 Cepheid variables; other 8 measurements are comprised within $0.1791 \leq z \leq 1.037$ as presented in Moresco et al.~\cite{Moresco2012}. These last were obtained from the differential spectroscopic evolution of early-type galaxies with respect to $z$. In particular, we use their $H(z)$ values computed assuming the BC03 model for the differential evolution. The last datum is an estimate of $H(z)r_d / (1+z)$ at $z = 2.3$, and it is derived from the transmitted flux fraction in the Ly$\alpha$ forest of over 48,000 quasars combined with CMB observations as showed by Busca et al.~\cite{Busca2013}. 

Assuming that $H(z)$ follows a Gaussian distribution and given that the error of each measurement is independent, we have that the likelihood is 
\begin{equation}\label{eq:lik_hz}
-2 \ln L_{H} = \sum_{i = 1}^{20} \frac{\left(H(z_i) - H^{\text{obs}}_i\right)^2}{\sigma_i^2} + \frac{\left.\left[\frac{H(z)r_d}{1+z} - H_{r}^{\text{obs}}\right]^2\right\vert_{z=z_{21}}}{\sigma_{21}^2},
\end{equation}
where $H_i^{\text{obs}}$ and $\sigma_i$ are the data points and their respective errors, $H(z) = H_0 E(z)$ and $E(z)$ is given by eq.~\eqref{eq:Ez}. The observed values of the $20$ measures of the Hubble function are depicted in table~\ref{tab:Hz} in appendix~\ref{app:sampling}. The last observable $H^{\text{obs}}_r$ is described in the footnote of the same table.

\section{Methodology and Validation}
\label{sec:meth_valid}

The observational data set represent just one realization of their underlying data probability distributions. So by construction, we cannot access the bias of an estimated function, e.g., $\hat{q}(z)$, by fitting it using this data set. As we discussed in section~\ref{sec:bv_trade}, the $\hat{q}(z)$ bias is inferred knowing both the expected value and the true value of $q(z)$ [see eq.~\eqref{eq:mse_2}]. This last is the missing piece to compute the bias-variance trade-off. However, one can indirectly infer the bias without knowing the underlying $q(z)$.\footnote {There are methods to infer the bias without the knowledge of the true underlying model, e.g., bootstrap~\cite{Efron1994}. In this work we do not explore these approaches, since their application is not straightforward for correlated data.} First, one fits the model obtaining the best fit for the real data. Then, using this best fit as the fiducial model, several mock catalogs are generated and, following the description in section~\ref{sec:bv_trade}, one calculates the bias and, consequently, the bias-variance trade-off. 

The problem with this procedure is that one must choose the parameters of the reconstruction method to compute the best fit. In particular, we have to fix the number of knots and the complexity of the function $\hat{q}(z)$, through the penalization parameter $\rel$. Thus, if the initial best fit for a given $\rel$ is already significantly biased, so will be the bias-variance trade-off analysis. In other words, performing a MC analysis of the bias-variance trade-off around the best fit does not take into account the variance of the estimated curve. Therefore, a better approach consists in using not only the best fit curve but a set of curves inside some statistical significance, i.e., curves whose parameters are inside some confidence interval of the best fit. This means that the procedure should be capable of reconstructing not only the best fit curve but also every other curves inside some significance level.

There is a high computational cost to study the bias-variance trade-off for a given fiducial curve. It is necessary to resample from the model $m$ times and to find the best fit for each realization. The whole calculation must be performed for different values of $\rel$ until the best bias-variance trade-off is attained. 

In this work, we address the problem by performing the MC analyses for three different fiducial curves and seven $\rel$ values (see sections~\ref{sec:mc_sneia} and \ref{sec:mc_alldata}). The functional forms of these fiducial models were purposely defined to have quite different features as shown in figure~\ref{fig:fiducial_models}. Given these distinct scenarios, we can explore the capability and efficiency of the proposed method in reconstructing $q(z)$ and also verify any dependence on the underlying model. Essentially, we want to find the $\rel$ value which best reconstructs all the three fiducial models.

We carry out this study considering: (i) only the SN Ia data and (ii) jointly the SN Ia, BAO and $H(z)$ measurements, in sections~\ref{sec:mc_sneia} and \ref{sec:mc_alldata}, respectively. Therefore, each realization is a (pseudo)randomly-generated catalog of SNe Ia $\{m_{B, a}, X_a, \mathcal{C}_a\}$ or random samples of all three observables, i.e., $\{\{m_{B,a}, X_a, \mathcal{C}_a\}, \{(D_V/r_d)_b\}, \{H_c\}\}$, where $a = 1,...,740$, $b = 1,...,6$ and $c = 1,...,21$. The methodology and algorithms to generate these samples are described in appendix~\ref{app:sampling}.

We use the Monte Carlo object of the \textit{Numerical Cosmology} Library (NumCosmo)~\cite{DiasPintoVitenti2014}, called \textsf{NcmFitMC}. This object proceeds as describe in algorithm~\ref{algo:NcmFitMC}. The code generates a minimum of $\mathsf{prerun}$ mock samples and the catalog of the best fit for each sample. 

In the list below, we summarize the steps to obtain $\hat{q}(z)$ using the MC method:
\begin{enumerate}
\item Define the redshift domain $\mathrm{D} = [z_{min}, z_{max}]$, which is determined by the real data sample as described in appendix~\ref{app:sampling}.
\item Define the fiducial model $q^\f(z)$, where $z \in \mathrm{D}$.
\item Choose a $\rel$ value and the number of knots $n$, which will be the same for all $\rel$ values to be tested. 
\item Run the MC algorithm \textsf{NcmFitMC} for $m = \mathsf{prerun}$ minimum realizations.
\item Finally, compute the MSE.
\end{enumerate}
These steps are repeated for different values of $\rel$ and their respective results are compared in order to determine the best scenario which minimizes the MSE [eq.~\eqref{eq:mse_2}] for a given $m_b$ [eq.~\eqref{eq:m_b}]. 

\begin{algorithm}
 \caption{\textsf{NcmFitMC} object implemented in NumCosmo.\label{algo:NcmFitMC}}
 \KwIn{$\vec{\theta}^\f \longrightarrow$ fiducial model.}
 \KwIn{$L_P(\vec{D},\vec{\theta}) \longrightarrow$ data set likelihood.}
 \KwIn{$\mathsf{prerun} \longrightarrow$ minimum number of samples.}
 \KwIn{$\mathsf{lre} \longrightarrow$ largest relative error.}
 \KwResult{$\vec{\theta}^{(l)}$ best fit catalog.}
 $l = 0$ \\
 \Repeat {$l \geq \mathsf{prerun}$ and $\mathsf{lre}^{(l)} < \mathsf{lre}$}
  {Generate catalog $\vec{D}^{(l)}$ from the fiducial model $\vec{\theta}^\f$.\\
   Find the best fit $\vec{\theta}^{(l)}$ for $\vec{D}^{(l)}$.\\
   Update the sample estimates of $\langle\theta_i\rangle$ and $\text{Cov}(\theta_i,\theta_j)$.\\
   Estimate the mean standard deviation $\sigma_{\langle\theta_i\rangle} = \sqrt{\text{Var}(\theta_i)/(l+1)}$. \\
   Calculate the largest relative error on the mean, $\mathsf{lre}^{(l)} = \max(\{\sigma_{\langle\theta_i\rangle}/\langle\theta_i\rangle\})$. \\
   Store the best fit $\vec{\theta}^{(l)}$ and the value of the likelihood $-2\ln(L_P(\vec{D}^{(l)},\vec{\theta}^{l}))$.\\
   $l = l + 1$.}
\end{algorithm}

\subsection{Monte Carlo analyses: SNe Ia}
\label{sec:mc_sneia}

\begin{figure}
\begin{center}
\includegraphics[scale=0.5]{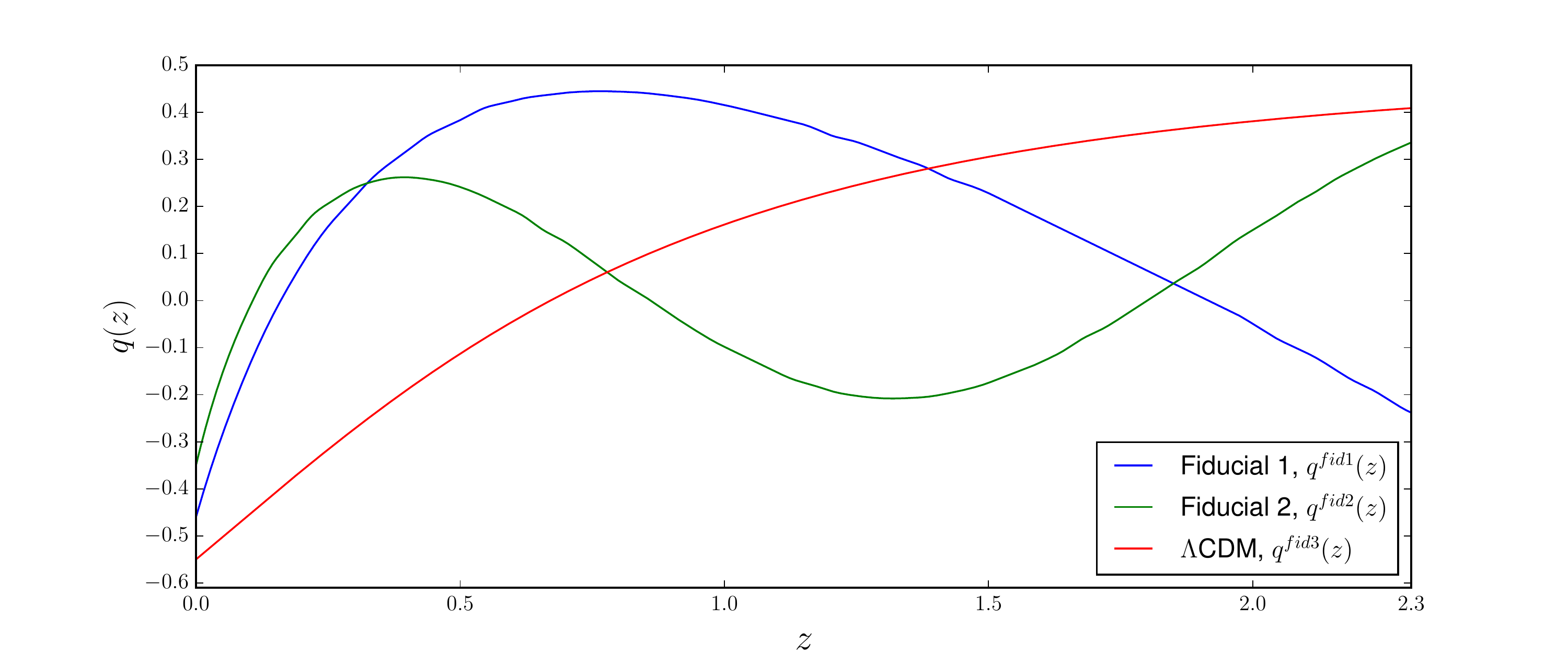}
\caption{The three $q^\f(z)$ fiducial models for which we study the reconstruction method via cubic spline. }
\label{fig:fiducial_models}
\end{center}
\end{figure}

We carry out the first analysis using the SNe Ia data, such that the realizations are generated using the covariance matrix and redshifts of the JLA sample (see appendix~\ref{app:sampling}). Thus, $\mathrm{D} = [0.0, 1.3]$ which we divide in 7 equally spaced intervals, i.e., $n + 1 = 8$ knots.\footnote{Following the discussion presented in the end of section~\ref{sec:spline}, we also performed the analyses using $n +1 = 6$ and $n+1 = 10$ knots. The results showed a small improvement from 6 to 8 knots, and a negligible variation from 8 to 10 knots.} The first and second fiducial models, denoted as $q^{\text{fid1}}(z)$ and $q^{\text{fid2}}(z)$, are defined by the blue and green curves, respectively, in figure~\ref{fig:fiducial_models}. The third one, $q^{\text{fid3}}(z)$, is the $\Lambda$CDM model (red curve), in which the cold dark matter and baryon density parameters are $\Omega_c = 0.3$ and $\Omega_b = 0.05$ and the DE EoS is $w = -1.0$. In all scenarios we assume a flat universe ($\Omega_k = 0$).\footnote{The densities $\Omega_c$ and $\Omega_b$ are the energy densities divided by the critical density today.} Finally, these three fiducial models are completely defined by fixing the SN Ia astrophysical parameters, namely $\alpha^\f = 0.141$, $\beta^\f = 3.101$, $M_1^\f = -19.05$ and $M_2^\f = -19.12$. These values correspond to the best-fit obtained in \cite{Betoule2014} considering $\Lambda$CDM and SNe Ia data (including both systematic and statistical errors).\footnote{In the reference, the authors use a different parametrization $(M_B, \Delta_M)$ such that $M_1 = M_B$ and $M_2 = M_B + \Delta_M$. } 

\begin{figure}
\begin{center}
\includegraphics[scale=0.41]{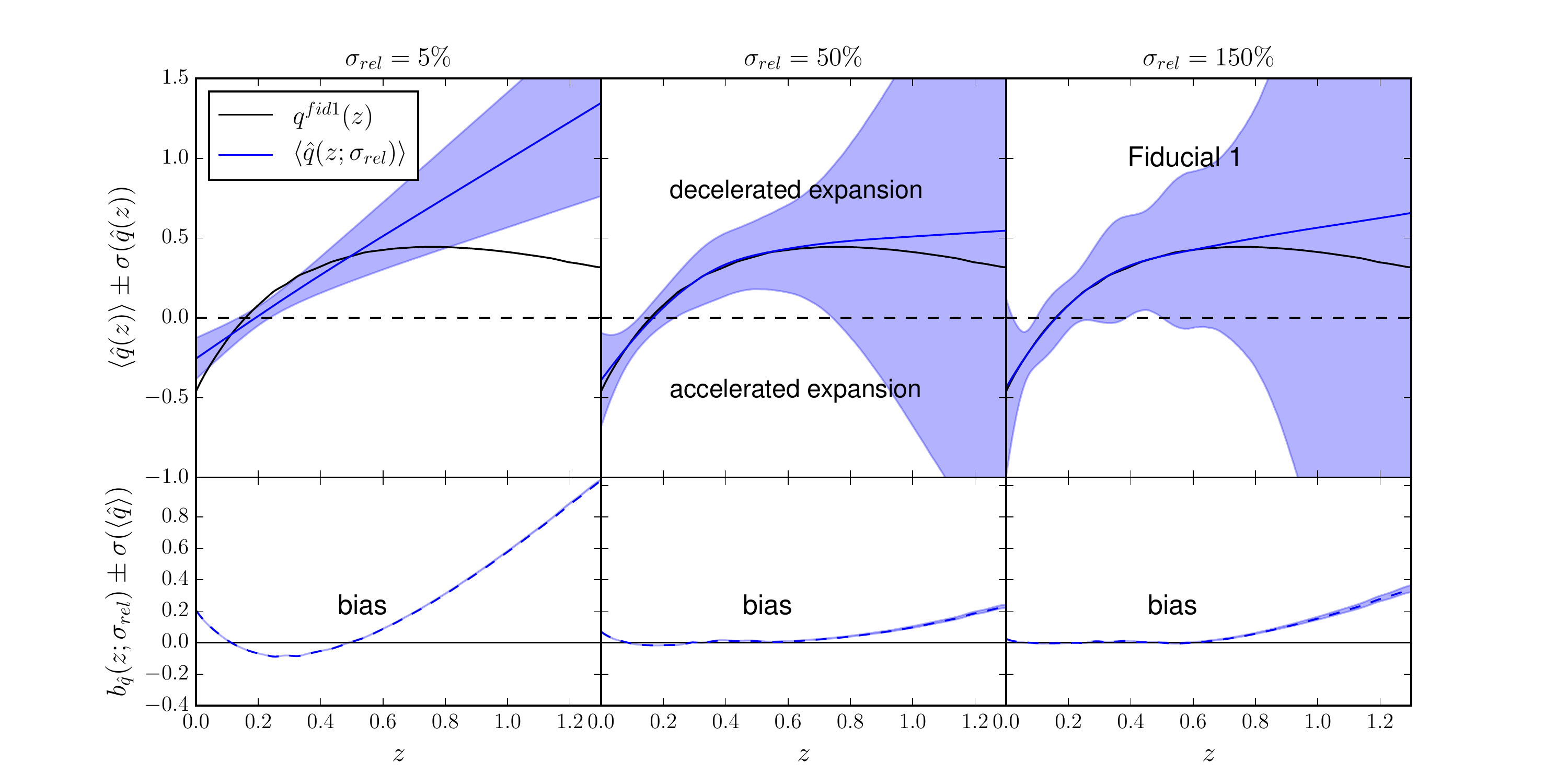}
\includegraphics[scale=0.41]{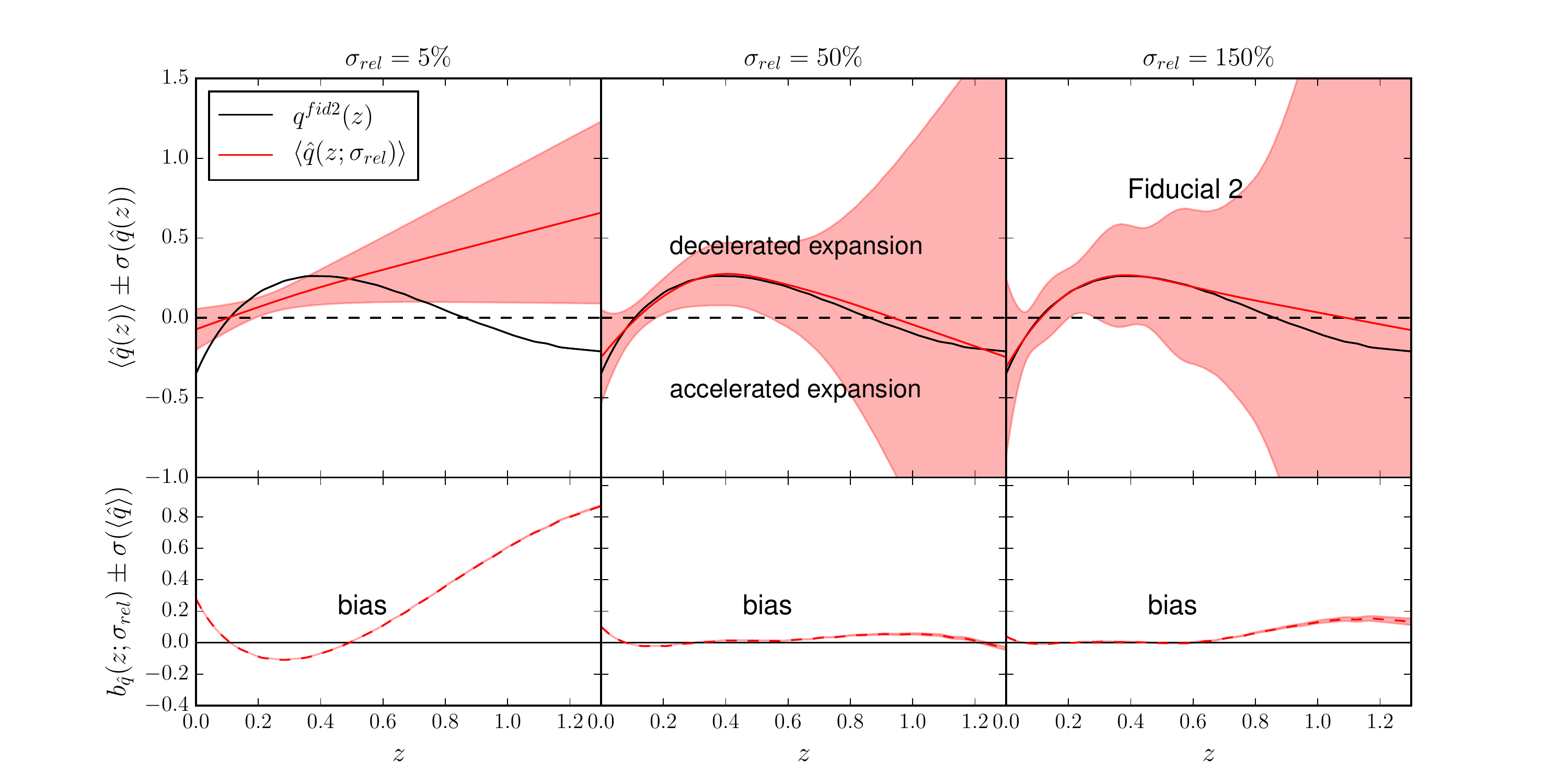}
\includegraphics[scale=0.41]{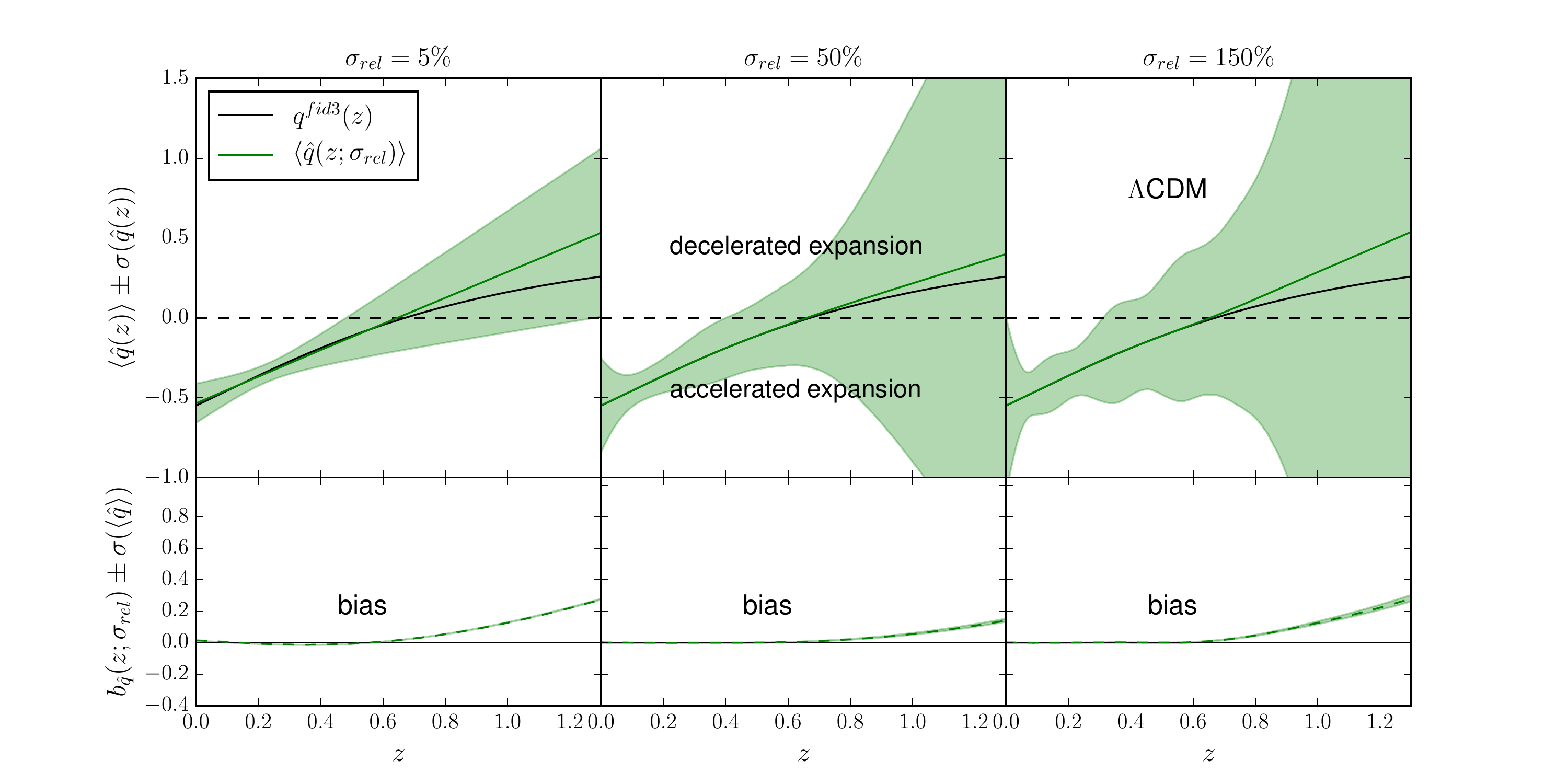}
\caption{The top part of each panel shows the reconstructed curve of $q(z)$ using 8 knots, MC approach and sampling from SNe Ia data. The colored (blue, red and green) lines and shaded regions are the mean function $\langle\hat{q}(z)\rangle$ and their $1\sigma$ error bar, respectively, obtained for a given $\rel$ and $q^\f(z)$. The black lines correspond to $q^{\f1}$ (upper panel), $q^{\f2}$ (middle) and $\Lambda$CDM (lower). The bottom part of each panel shows the bias (dashed lines) and its $1\sigma$ error bar of the mean curve.}
\label{fig:qz_qs8}
\end{center}
\end{figure}

In order to obtain the features of the reconstruction procedure as a function of the penalty factor (through $\rel$) and $q^\f(z)$, we perform the MC analyses considering 7 different $\rel$ values,
$\rel = \{5\%, 15\%, 30\%, 50\%, 75\%, 100\%, 150\% \},$ for each fiducial model. Independently of $\rel$ and $q^\f(z)$, we standardize our analyses fixing the number of realizations to $m= 42000$, with which we obtain small $\mathsf{lre} < 1\%$ (algorithm~\ref{algo:NcmFitMC}) in all cases. Then, for each mock catalog, we minimize the function
\begin{equation}\label{eq:f_snia}
-2 \ln (L_{\text{SNIa},P}) = -2 \ln (L_{\text{SNIa}}) + \sum_{i = 2}^{n-1} P_i(\rel),
\end{equation}
with respect to the free parameters 
\begin{equation}\label{eq:set_p1}
\vec{\theta} \doteq \{\hat{q}_0, \, \hat{q}_1, \, \hat{q}_2, \, \hat{q}_3, \, \hat{q}_4, \, \hat{q}_5, \, \hat{q}_6, \, \hat{q}_7, \, \alpha, \, \beta, \, M_1, \, M_2\},
\end{equation} where the right-side terms of eq.~\eqref{eq:f_snia} are given in eqs.~\eqref{eq:lik_snia} and \eqref{eq:penalty_fac}, respectively. Therefore, the expected values $\langle \hat{q}(z; \rel) \rangle$ and the error bars $\sigma(\hat{q}(z; \rel)) \equiv \sqrt{\text{Var}(\hat{q}(z; \rel))}$ of the reconstructed function $\hat{q}(z; \rel)$ are estimated by computing eqs.~\eqref{eq:expec_val} and \eqref{eq:cov_q}.

Figure~\ref{fig:qz_qs8} displays the reconstructed curves (colored solid lines) for $\rel = 5\%, 50\%$ and $150\%$, along with their respective fiducial models (black lines). As we discussed in section~\ref{sec:penalty}, a small $\rel$ implies in a big constraint on the function complexity. Indeed, we see that $\rel = 5\%$ imposes $\hat{q}(z; 5\%)$ to be a linear function independently of the underlying fiducial model (upper-left part of the three panels).  

Naturally, if the true model differs from a linear function, the result will be strongly biased and, consequently, the estimated curve will not be capable to recover the true form. In particular, $\hat{q}(z; 5\%)$ is outside the $1\sigma$ error bar in a large fraction of the redshift interval for $q^{\f1}(z)$ and $q^{\f2}(z)$. The bottom-left part of the three panels in figure~\ref{fig:qz_qs8} show the bias (colored dashed lines) as a function of the redshift and its $1\sigma$ error bar.

Overall, until $z \lesssim 0.6$, the bias $b(z; \rel)$ decreases as the $\hat{q}(z; \rel)$ function complexity (i.e., $\rel$) increases, such that the reconstructed function approximates better and better the fiducial curve. On the other hand, due to the small amount of data at higher redshifts, mainly for $z \gtrsim 1.0$, the standard deviation $\sigma(\hat{q}(z; \rel))$ greatly increases, as evinced in figure~\ref{fig:qz_qs8} by the colored shaded areas. As a direct result, we have that the constraints on the highest parameters $\hat{q}_i$ are degenerated causing an increment on $b(z; 150\%)$, for $z \gtrsim 0.6$, in comparison to $b(z; 50\%)$ for all three fiducial models, as shown in figure~\ref{fig:qz_qs8}.

\begin{figure}[h]
\begin{center}
\includegraphics[scale=0.41]{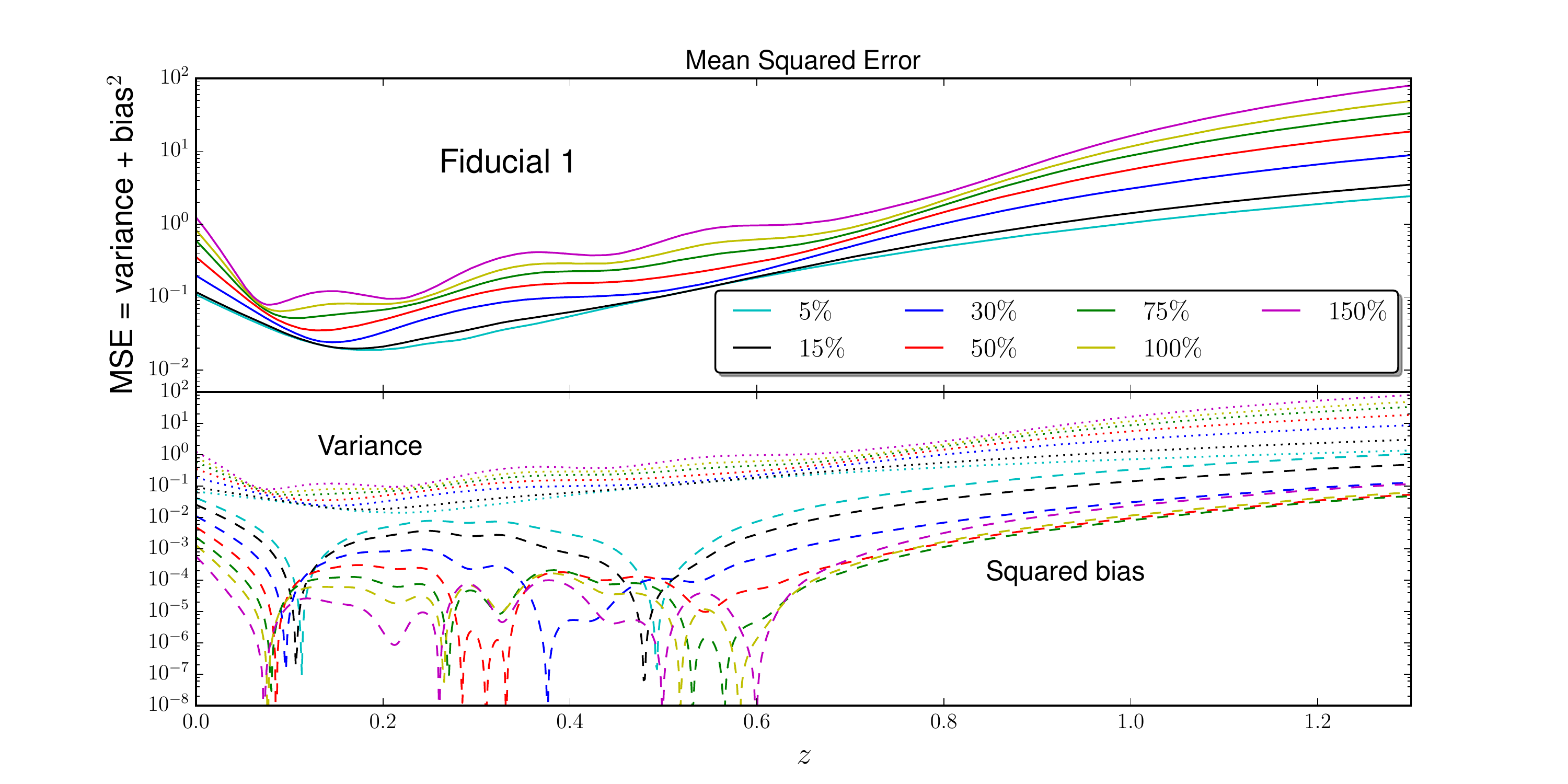}
\includegraphics[scale=0.41]{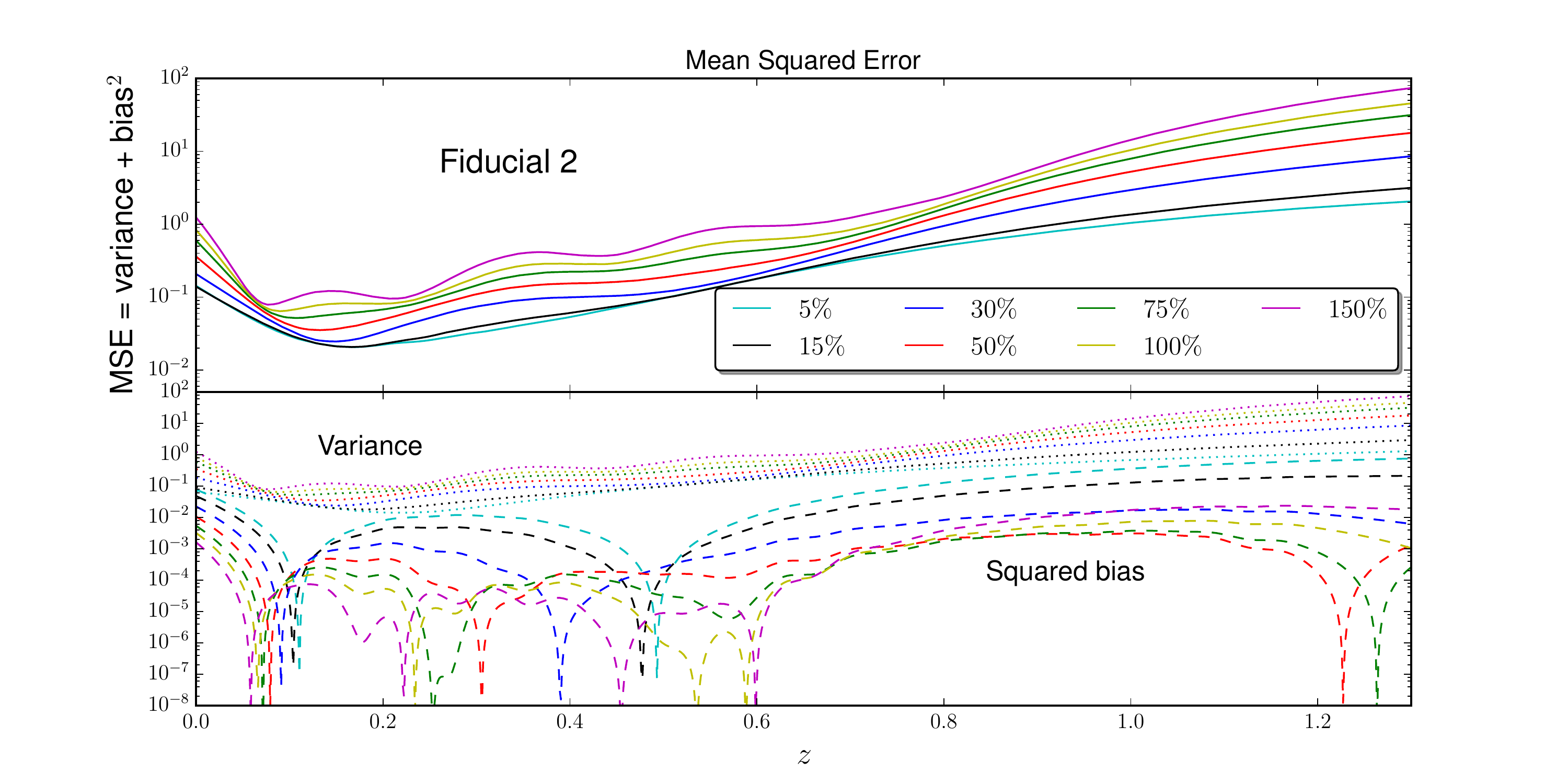}
\includegraphics[scale=0.41]{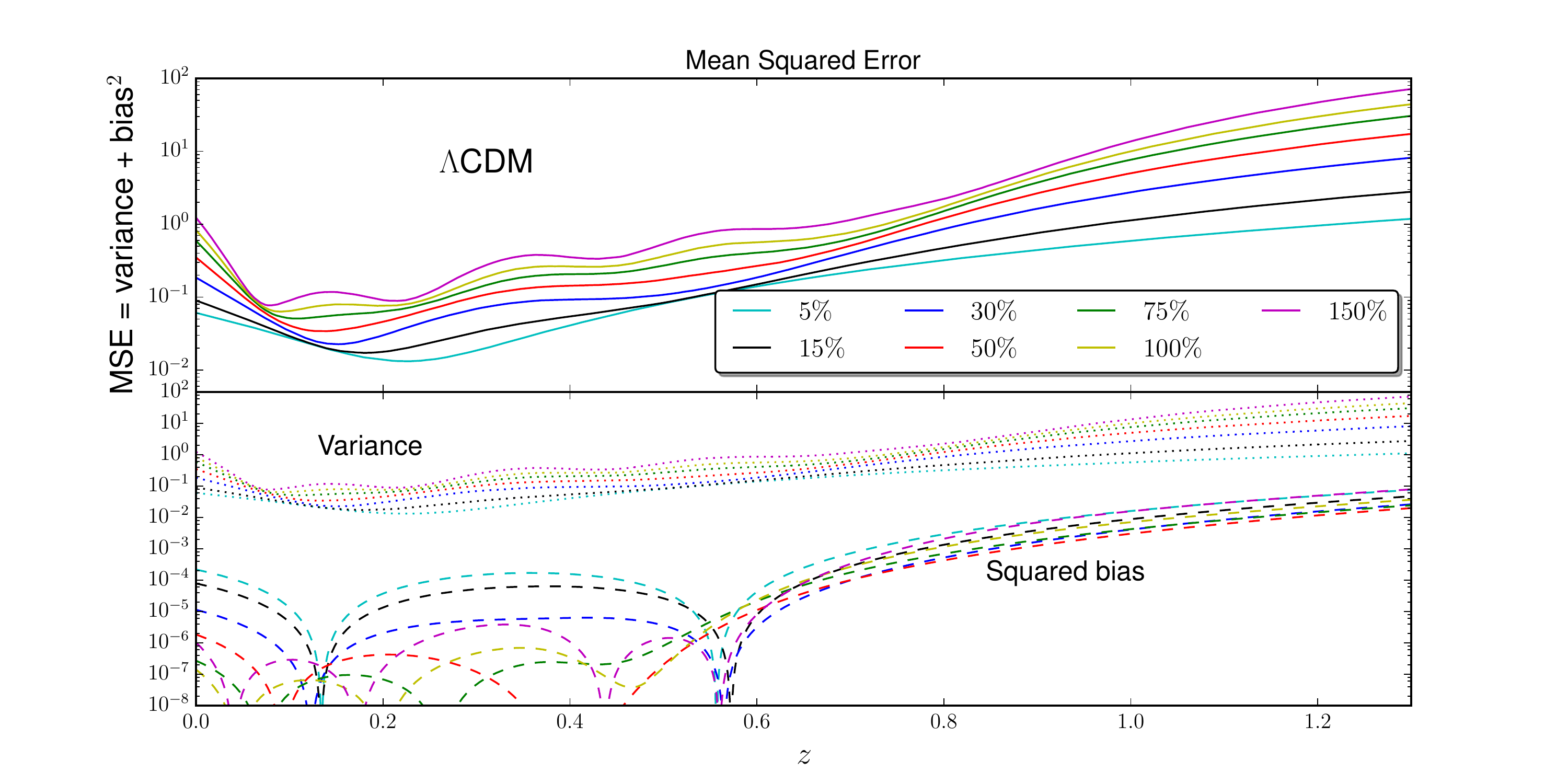}
\caption{The top part of each panel shows the MSE of the reconstructed $\hat{q}(z)$ for 7 different $\rel$ values $\in [5\%, 150\%]$, and obtained using 8 knots, MC approach and sampling from SNe Ia data. The upper, middle and lower panels refer to $q^{\f1}(z)$, $q^{\f2}(z)$ and $\Lambda$CDM models, respectively. The MSE decomposition into variance (dotted lines) and squared bias (dashed lines) is displayed in the bottom part of each panel.}
   \label{fig:mse_qs8}
\end{center}
\end{figure}

In order to define which $\rel$ provides the best reconstructed curve, we compute the MSE [eq.~\eqref{eq:mse_2}] as a function of $z$ of all 21 reconstructed curves. Figure~\ref{fig:mse_qs8} shows the MSE (solid lines) for each $\rel$ and $q^{\f1}(z)$ (upper panel), $q^{\f2}(z)$ (middle) and $\Lambda$CDM (lower) models. For these three fiducial models, $\hat{q}(z; 5\%)$ is the function with the smallest MSE for any $m_b$. However, these are also the most biased reconstructed curves and, even through visual inspection (figure~\ref{fig:qz_qs8}), they do not give satisfactory reconstructions to their respective true $q^\f(z)$. Nevertheless, if we have only analyzed the fiducial curve closer to the best fit one (fiducial 3), we would wrongly conclude that $\rel = 5\%$ provides the smallest MSE with a insignificant bias (see the last panel in figure~\ref{fig:mse_qs8}).

More importantly, when performing the analysis with real data, we will be able to estimate the error bars but not the bias. We can note in figure~\ref{fig:mse_qs8} that the MSE for $\rel = 5\%$ has about the same contribution from bias and variance. Thus, if we choose the smallest MSE ($\rel = 5\%$) disregarding $m_b$, the estimated error in the real data analysis would provide only half of the total uncertainty in the reconstruction. In a conservative approach one would estimate the variance and then double by hand the error bars.\footnote{In a more careful analysis, one could calculate the bias for several fiducial models and their respective upper limits. Then, this value could be added to the curve obtained from the real data.} Notwithstanding, inspecting the $\rel = 5\%$ reconstructions in figure~\ref{fig:qz_qs8}, we note that this reconstruction looses all information about the shape of the curve. So even correcting the error bar for $\hat{q}(z)$ by doubling the computed error, for example, the form of the curve will always be close to a straight line.

The objective of this work is to reconstruct the form of the kinematic curve. We select the best reconstructed function (smallest MSE) requiring the bias to be at most $10\%$ of the total error, i.e., $m_b  = 0.1$. Making this imposition, the estimated variance for the fit using real data will provide a good approximation of the uncertainty of the reconstruction. Therefore, taking into account the three fiducial models, we find that the best bias-variance trade-off as a function of $z$ is achieved for $\rel = 30\%$. It is worth noting that the MC results for $\alpha$, $\beta$, $M_1$ and $M_2$ are unbiased (bias smaller than $0.1\%$) for all 21 cases that we studied.

\subsection{Monte Carlo analyses: SNe Ia + BAO + $H(z)$}
\label{sec:mc_alldata}

In this section, we perform the MC analyses combining the SNe Ia, $H(z)$ and BAO data. In this case, we equally divide the redshift interval, $\mathrm{D} = [0.0, 2.3]$, using $n + 1 = 12$ knots. The fiducial models and parameters are those defined in section~\ref{sec:mc_sneia} and, additionally, $r_d^\f = 103.5 \, h^{-1}\text{Mpc}$ and $H_0^\f = 73.0 \, \text{km} \, \text{s}^{-1} \text{Mpc}^{-1}$.
Similarly, the MC study is done for the same $\rel$ set and the three fiducial models. In view of the increased amount of data, the present $q(z)$ reconstruction is better constrained and $m= 20000$ mock catalogs (for each MC analysis) are sufficient to provide $\langle \hat{q}(z; \rel) \rangle$ with small $\mathsf{lre} < 1\%$ (see algorithm~\ref{algo:NcmFitMC}).  


Since the SN Ia, BAO and $H(z)$ data sets are independent, the joint likelihood is
\begin{equation}
-2\ln(L_{\text{SBH},P}) = -2 \left( \ln L_{\text{SNIa}} + \ln L_{\text{BAO}} + \ln L_{H} \right) + \sum_{i = 2}^{n-1} P_i(\rel),
\end{equation}
where $-2 \ln L_{SNIa}$, $-2 \ln L_{BAO}$ and $-2 \ln L_{H(z)}$ are given by equations~\eqref{eq:lik_snia}, \eqref{eq:lik_bao} and \eqref{eq:lik_hz}, respectively. Then, for each realization, we compute the best-fitting values of the following 18 parameters,
\begin{equation}\label{eq:set_p2}
\vec{\theta} \doteq \{\hat{q}_0, \, \hat{q}_1, \, \hat{q}_2, \, \hat{q}_3, \, \hat{q}_4, \, \hat{q}_5, \, \hat{q}_6, \, \hat{q}_7, \, \hat{q}_8, \, \hat{q}_9, \, \hat{q}_{10}, \, \hat{q}_{11}, \, \alpha, \, \beta, \, M_1, \, M_2, H_0, r_d\},
\end{equation}
and, with them, we calculate the expected value of each parameter estimator, $\langle \vec{\theta} \rangle$, and their covariance matrix.

\begin{figure}[h]
\begin{center}
\includegraphics[scale=0.40]{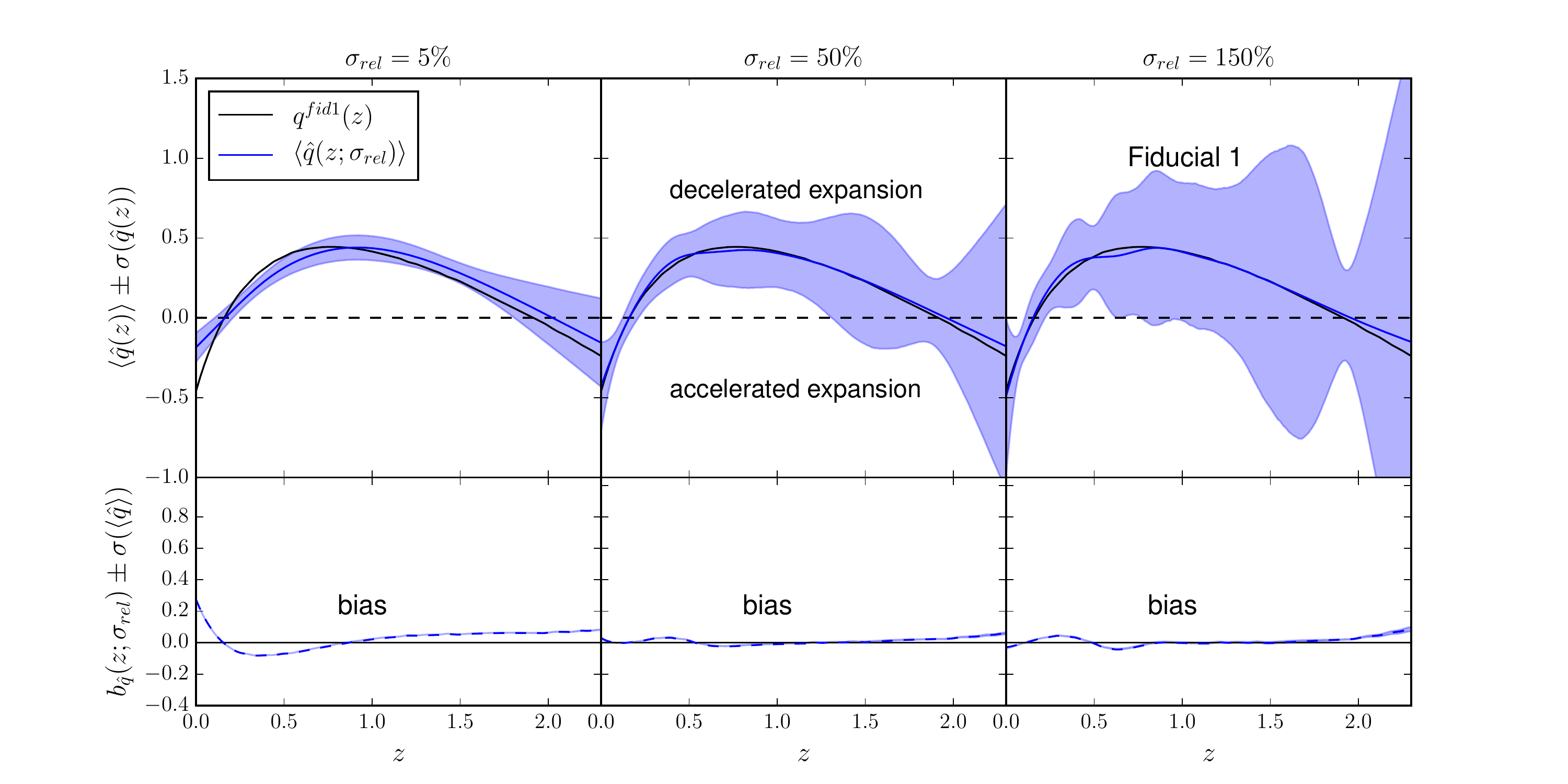}
\includegraphics[scale=0.40]{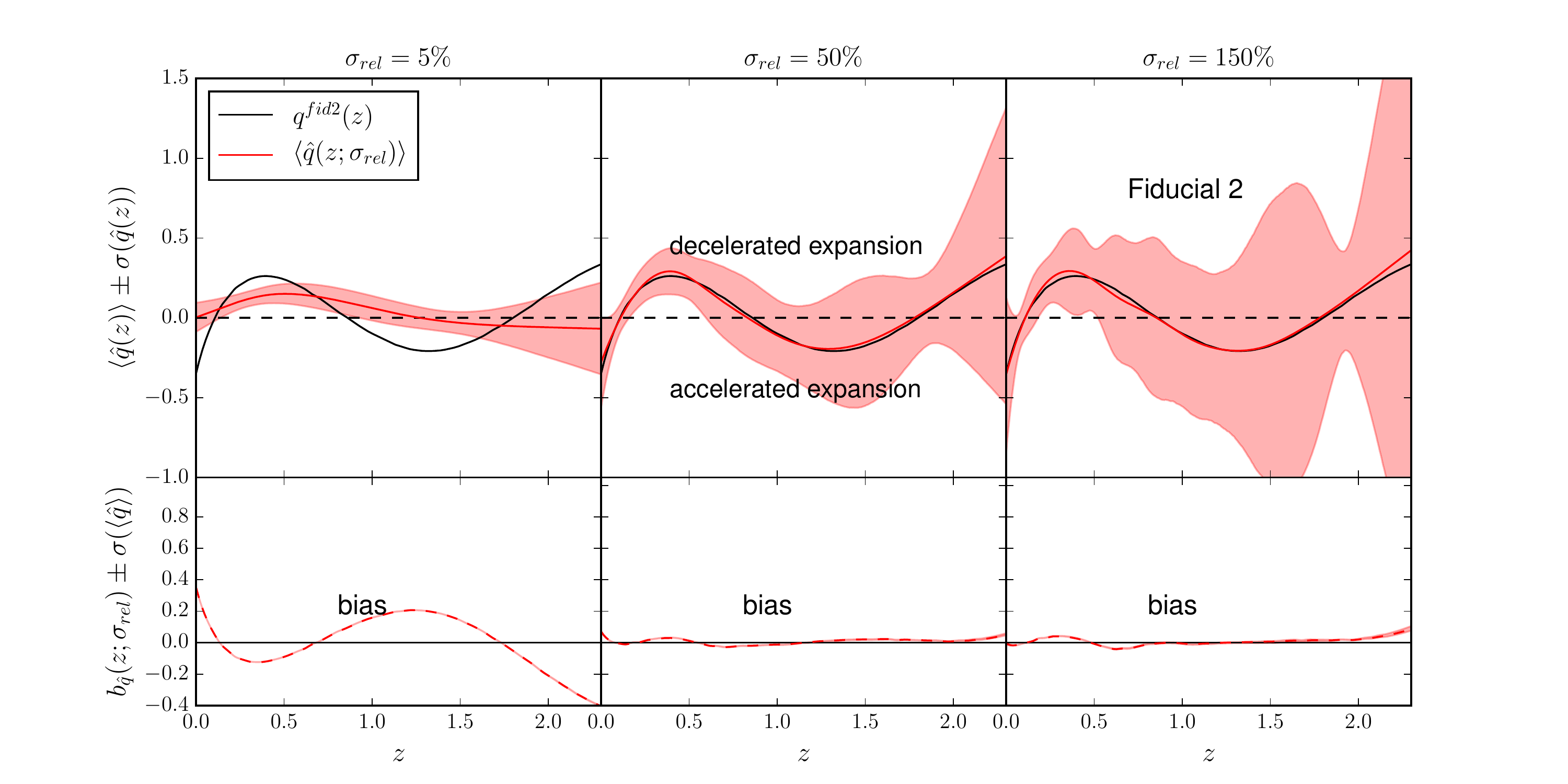}
\includegraphics[scale=0.40]{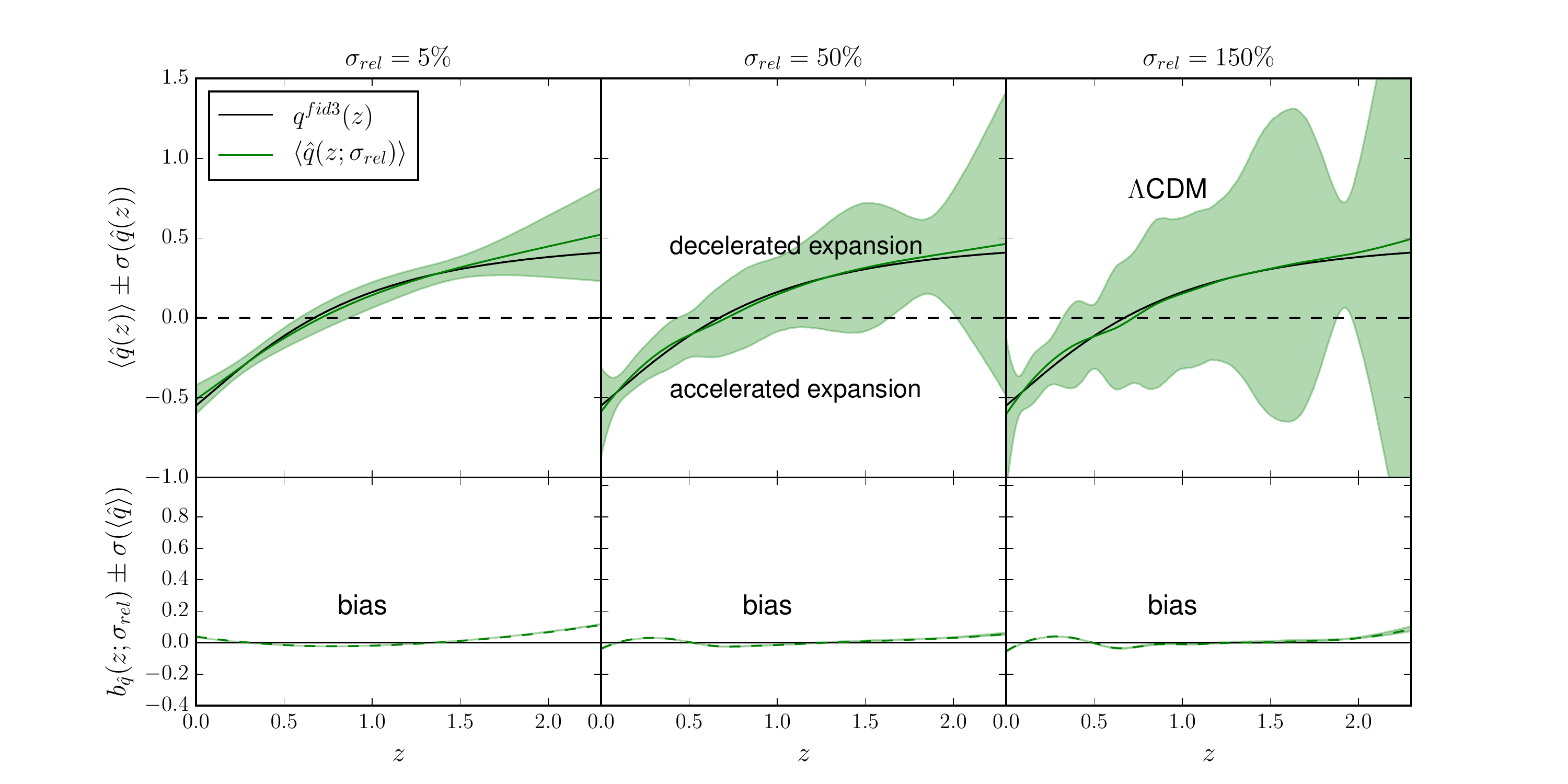}
\caption{The top part of each panel shows the reconstructed curve of $q(z)$ using 12 knots, MC approach and sampling from SNe Ia, BAO and $H(z)$ data. The colored (blue, red and green) lines and shadow regions are the mean function $\langle\hat{q}(z)\rangle$ and their $1\sigma$ error bar, respectively, obtained for a given $\rel$ and $q^\f(z)$. The black lines correspond to $q^{\f1}$ (upper panel), $q^{\f2}$ (middle) and $\Lambda$CDM (lower). The bottom part of each panel shows the bias (dashed lines) and its $1\sigma$ error bar of the mean curve.}
   \label{fig:qz_qs12}
\end{center}
\end{figure}

\begin{figure}[h]
\begin{center}
\includegraphics[scale=0.40]{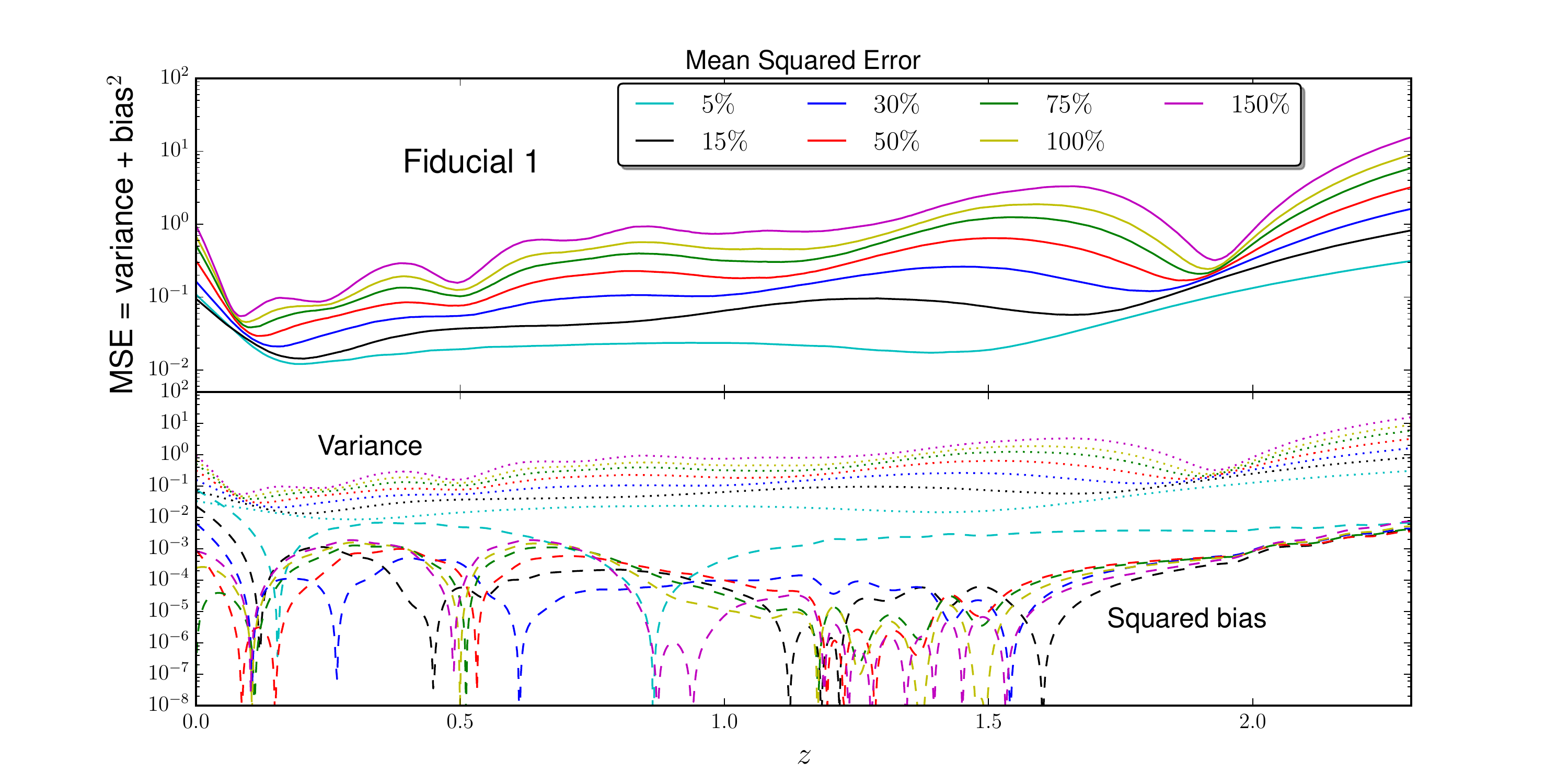}
\includegraphics[scale=0.40]{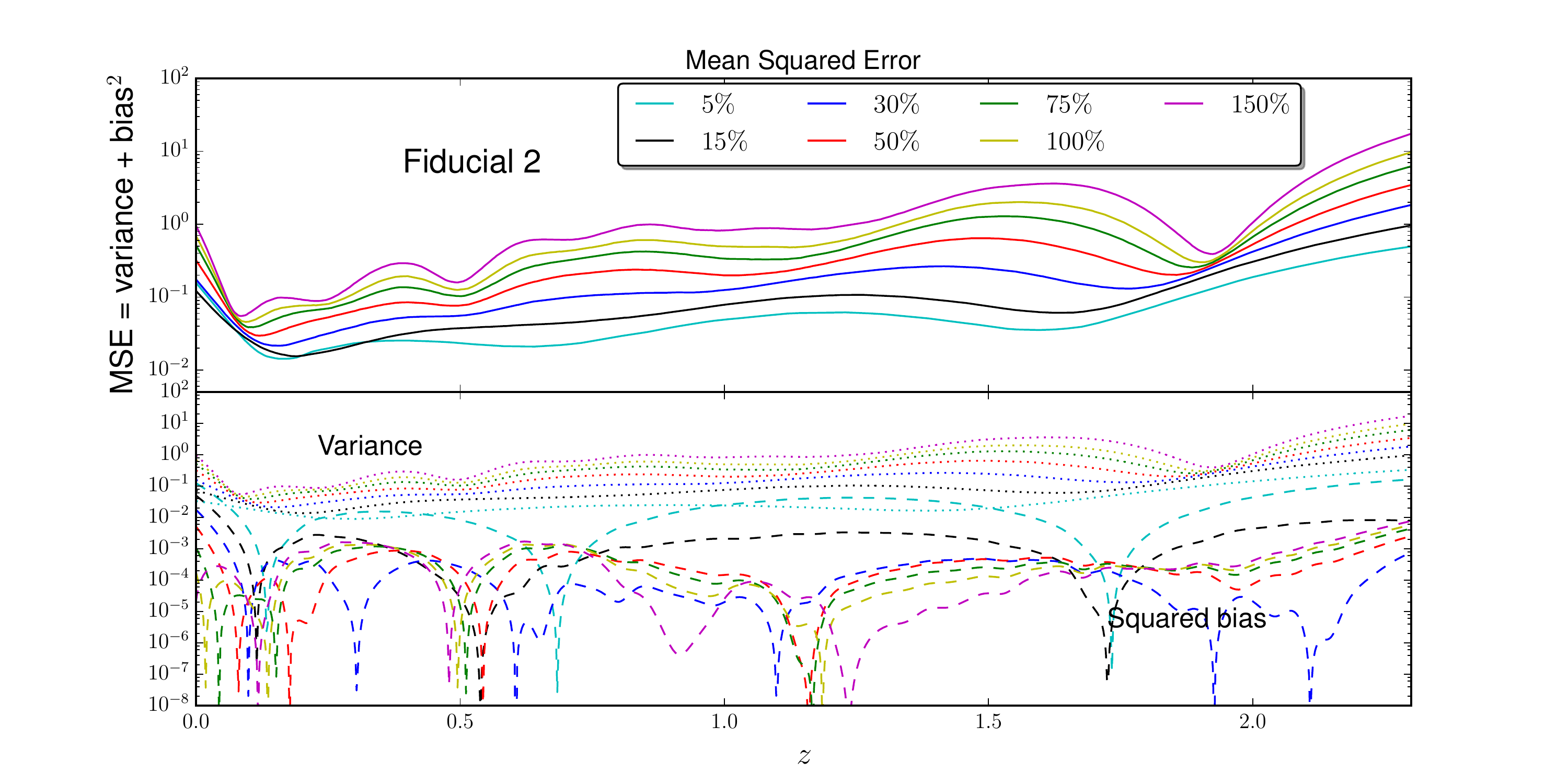}
\includegraphics[scale=0.40]{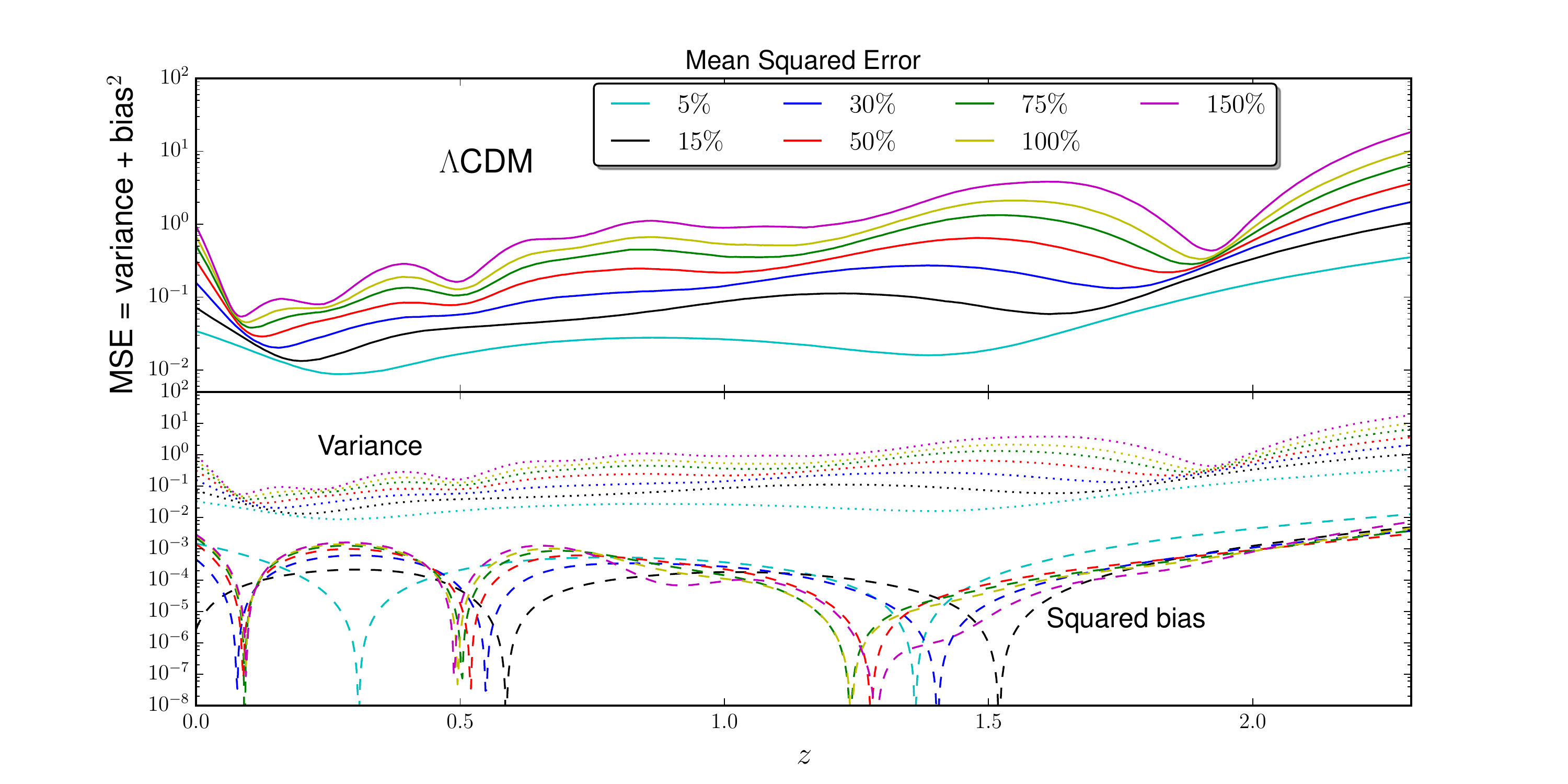}
\caption{The top part of each panel shows the MSE of the reconstructed $\hat{q}(z)$ for 7 different $\rel$ values $\in [5\%, 150\%]$, and obtained using 12 knots, MC approach and sampling from SNe Ia, BAO and $H(z)$ data. The upper, middle and lower panels refer to $q^{\f1}(z)$, $q^{\f2}(z)$ and $\Lambda$CDM models, respectively. The MSE decomposition into variance (dotted lines) and squared bias (dashed lines) is displayed in the bottom part of each panel.}
   \label{fig:mse_qs12}
\end{center}
\end{figure}

Analogously to the results presented in section~\ref{sec:mc_sneia}, in figure~\ref{fig:qz_qs12} we show the reconstructed curves, $\langle \hat{q} (z; \rel) \rangle$ (colored solid lines), the biases (dashed lines) and their respective $1\sigma$ error bars (shaded areas) for $\rel = 5\%, 50\%$ and $150\%$ and the three fiducial models. As expected, combining SN Ia, BAO and $H(z)$ data decreases $\sigma(\hat{q}(z; \rel))$ and we note, in the three upper-left panels of figure~\ref{fig:qz_qs12}, that these improved constraints no longer restrict $\langle \hat{q}(z; 5\%) \rangle$ to be a linear function.\footnote{This is natural since, in this case, we have more knots and we would need a smaller $\rel$ value to constrain into a linear curve.} Besides, as displayed in figure~\ref{fig:mse_qs12}, it also leads to a stronger dependency of $b(z; \rel)$ on the fiducial model. 

For example, the squared bias function of $q^{\f2}(z)$ (dashed lines on the middle panel of figure~\ref{fig:mse_qs12}) presents a large variation with respect to $\rel$, similar to what we obtained considering only SN Ia data. On the other hand, the bias function of the $\Lambda$CDM fiducial model is nearly invariant with respect to $\rel$ (lower panel of figure~\ref{fig:mse_qs12}). This highlights that, giving a sufficient amount of data, if the true model has low complexity form then, naturally, a satisfactory reconstruction will rapidly be reached even for small $\rel$. In particular, taking into account the bias-variance trade-off and $m_b = 0.1$, the best reconstructed curve for the $\Lambda$CDM fiducial model is obtained for $\rel = 15\%$.

As it was shown in figures~\ref{fig:mse_qs8} and \ref{fig:mse_qs12}, the MSE is dominated by the variance of the fitted parameters and, consequently, by the measurement errors of the observable data sets. In this way, we note that the MSE is independent of the fiducial model, depending only on the penalty factor, i.e., $\rel$. This aspect points out the fact that an statistical inference with relative small error bars can be completely misleading. For example, the well-constrained function $\hat{q}(z; 5\%)$, obtained for $q^{\f2}(z)$, is not at all a good reconstruction of the true model, diverging with more than $1\sigma$ in a large fraction of the redshift interval. Ultimately, following the discussion on section~\ref{sec:PM}, the nature of the biases for $\rel = 5\%$ and $150\%$ can be classified as form-bias and estimator-bias, respectively.

As in section~\ref{sec:mc_sneia}, we expect that a good reconstruction is one which does not present a significant bias. Therefore, applying the same requirement $m_b = 0.1$, we obtain that $\rel = 30\%$ provides the best balance between bias and variance, given the current data sets, to reconstruct the $q(z)$ curve. 

In summary, the MC outcomes (sections~\ref{sec:mc_sneia} and \ref{sec:mc_alldata}) show that our method is efficient in reconstructing $q(z)$ and, given the current errors of the observational data, $\rel = 30\%$ is a safe and conservative choice to recover $q(z)$ imposing minimal assumptions and guaranteeing that the reconstructed curve will not be bias dominated.

\section{Results}
\label{sec:real_reconst}

Defined the best estimator, $\hat{q}(z; 30\%)$, we now obtain the deceleration function given the real JLA, BAO and $H(z)$ samples (see section~\ref{sec:data}). As before, we reconstruct $q(z)$ (i) in the redshift interval $\mathrm{D} = [0, 1.3]$ using JLA SN Ia data and (ii) in $\mathrm{D} = [0, 2.3]$ combining those three observable data. 

The likelihood function is now interpreted as the posterior distribution $$\mathcal{P}(\vec{\theta} | \vec{D}) = L_P(\vec{D}| \vec{\theta}),$$ since we are assuming flat priors on all parameters. We then use the NumCosmo algorithm \textsf{NcmFitESMCMC},\footnote{The algorithm is describe in~\cite{Goodman2010} and was implemented in C in the NumCosmo library~\cite{DiasPintoVitenti2014}. Another unrelated implementation in Python is described in~\cite{Foreman-Mackey2013}.} which implements an ensemble sampler with affine invariance for Markov Chain Monte Carlo analysis, to compute the mean function $\overline{q}(z)$ given the JLA sample and its $68.27\%$, $95.45\%$ and $99.73\%$ confidence intervals (CI).  We ran 100 chains, computing $8\times 10^5$ points in the 12-dimensional parametric space, shown in eq.~\eqref{eq:set_p1}, attaining a multivariate potential scale reduction factor  (MPSRF) equal to $1.016$.\footnote{The MPSRF is a diagnose tool used to check the convergence of a Markov Chain, which was originally proposed in~\cite{Brooks1998}. A value $<1.2$ indicates the convergence.} The best-fitting values and the error bars of the SNe Ia parameters are $\alpha =0.141 \pm 0.007$, $\beta = 3.108 \pm 0.081$, $M_1 = -19.05 \pm 0.03$ and $M_2 = -19.12 \pm 0.03$.

\begin{figure}[h]
\begin{center}
\includegraphics[scale=0.5]{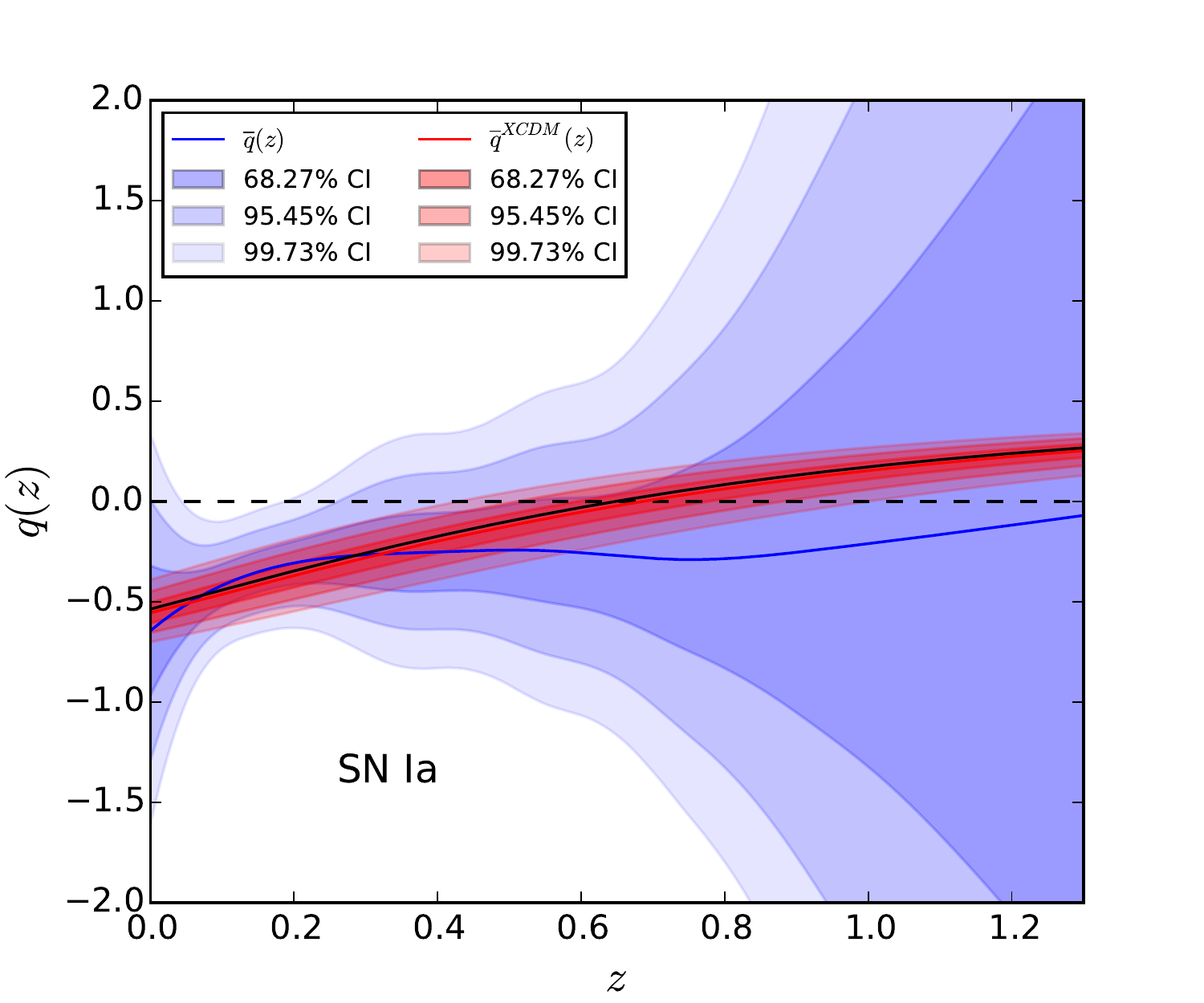}
\includegraphics[scale=0.5]{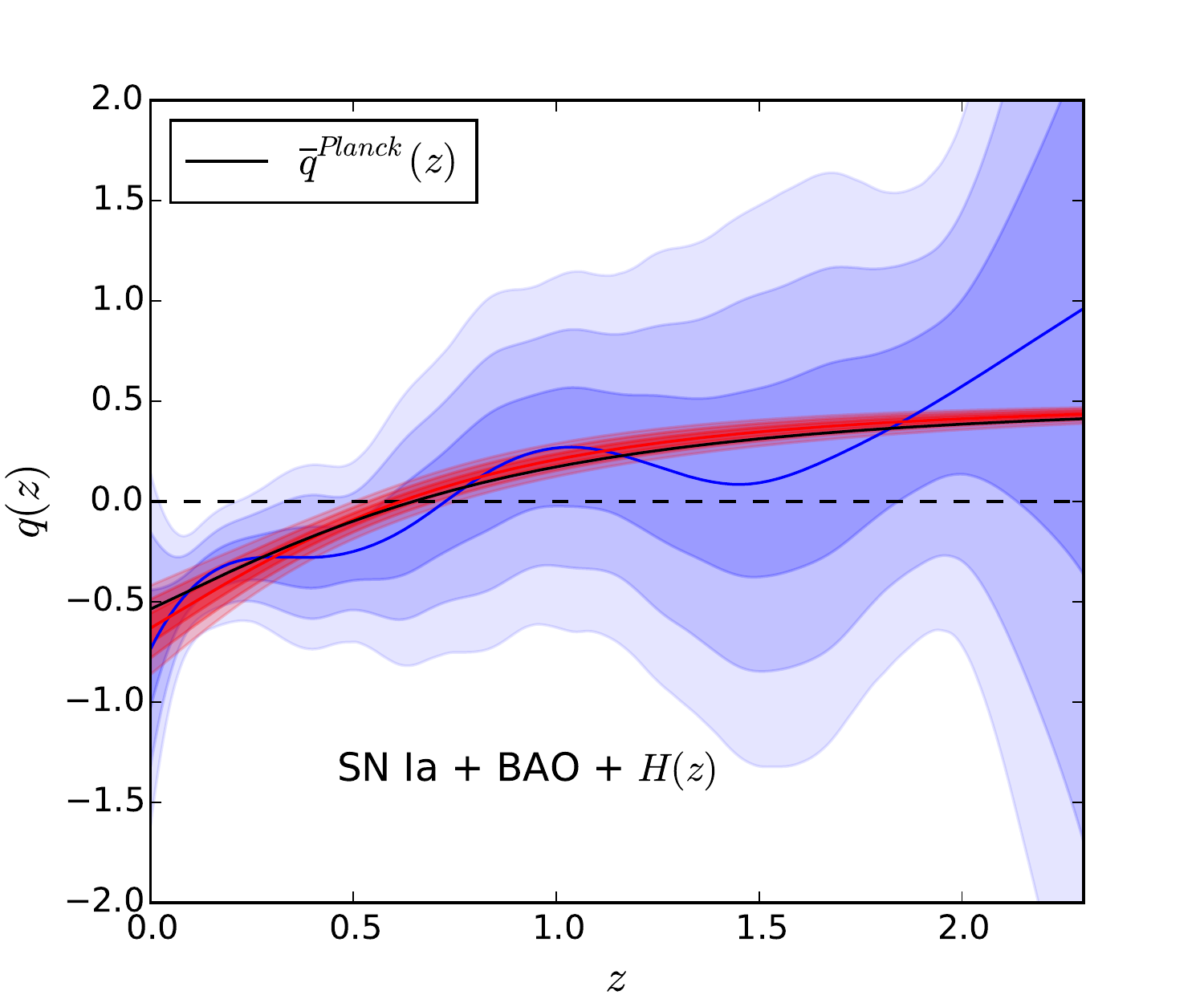}
\caption{The model-independent reconstructed $q(z)$ (blue solid line) and its $68.23\%$, $95.45\%$ and $99.73\%$ confidence intervals (blue shaded areas), for $\rel = 30\%$, using SNe Ia (left panel) and SNe Ia + BAO + $H(z)$ data (right panel). The red line and contours are the $q(z)$ mean and its CI's obtained, assuming XCDM model, fitting $\Omega_c$ (with $w = -1$) (left panel) and $(H_0, \Omega_c, w)$ (right panel) along with the SNe Ia nuisance parameters. The black lines correspond to the Planck+BAO+JLA+$H_0$ best-fit assuming $\Lambda$CDM model.}
   \label{fig:real_qz}
\end{center}
\end{figure}

The mean (blue line) and CI's (blue shaded areas) are displayed in the left panel of figure~\ref{fig:real_qz}, where we note that the ESMCMC $68.27\%$ CI is in agreement to the respective MC error bar. Naturally, as in the MC study, the difficulty in constraining $q(z)$ with minimal assumptions is that the result is highly degenerated. So, besides the fact that $\overline{q}(z) < 0$ in the entire redshift, this is not statistically significant. The main results of this $q(z)$ reconstruction, using only SNe Ia data, are: the indication of a transition redshift at $z_{\tiny{T}} \simeq 0.53$ with $68.27\%$ significance level, and the evidence of an accelerated expansion phase $\geq 99.73\%$ within the redshift interval $\sim [0.04, 0.19]$. 

To compare this result with the flat $\Lambda$CDM model, i.e., assuming GR and DE EoS given by $w = -1.0$, we carry out the ESMCMC in the 5-dimensional parametric space $$\vec{\theta} \doteq (\Omega_c, \alpha, \beta, M_1, M_2).$$ For this, we fixed the other cosmological parameters to the JLA best-fit. We obtain the following best-fitting values and standard deviations: $\Omega_c = 0.244 \pm 0.034$, $\alpha = 0.141 \pm 0.007$, $\beta = 3.103 \pm 0.081$, $M_1 = -19.05 \pm 0.023$ and $M_2 = -19.12 \pm 0.026$.
The mean $\overline{q}^{\text{XCDM}} (z)$ (red line) and the CI's (red shaded areas) are consistent with our model-independent reconstruction overall redshift range, as shown in the left panel of figure~\ref{fig:real_qz}. In both analyses, the SN Ia nuisance parameters $\alpha$ and $\beta$ are weakly (anti-) correlated ($\lesssim 0.1$) to $\{\hat{q}_i\}$ $(i = 0,..., 7)$ and $\Omega_c$. $M_1$ and $M_2$ are also weakly  correlated to most parameters excepted for $q_0$ and $\Omega_c$, in which there are moderate correlations $\sim 0.56 - 0.66$. At last, we also plot the Planck best-fitting curve (black line), $\overline{q}^{\text{Planck}}(z)$, which we computed using the Planck+BAO+JLA+$H_0$ best-fitting parameters obtained assuming $\Lambda$CDM \cite{Planck2015}, namely, $H_0 = 67.74$, $\Omega_c = 0.259$ and $\Omega_b =0.049$. We note that $\overline{q}^{\text{Planck}}(z)$ is inside the $68.27\%$ CI in the entire redshift interval. 

We follow the same procedure to compute $\overline{q}(z)$ and their CI in the redshift interval $[0.0, 2.3]$ given the JLA, BAO and $H(z)$ data. In this case, $8\times 10^5$ points are calculated in the 18-dimensional parametric space [eq.~\eqref{eq:set_p2}], getting $\text{MPSRF} = 1.033$. In this case, the best-fitting and standard deviation values of the non-$q_i$ parameters are $H_0 = 71.68 \pm 1.69 \, \text{km} \, \text{s}^{-1} \text{Mpc}^{-1}$, $r_d = 101.15 \pm 1.8 \, h^{-1}\text{Mpc}$, $\alpha = 0.141 \pm 0.007$, $\beta = 3.111 \pm 0.081$, $M_1 = -19.01 \pm 0.05$ and $M_2 = -19.07 \pm 0.05$. We note in the right panel of figure~\ref{fig:real_qz} an expressive improvement on the constraints due to the combined data, regarding that we are assuming only FLRW metric and flat space, and given the high dimension of the parametric space. In this case, we obtain $z_{\tiny{T}} \simeq 0.58$ with $68.27\%$ CI and we can attest with significance $\geq 99.73\%$ that the universe is accelerating within $0.02 \lesssim z \lesssim 0.22$. At last, we also have an indication of a decelerated phase in the redshift interval $1.84 \lesssim z \lesssim 2.13$ with $68.27\%$.

We compare the model-independent $q(z)$ reconstruction with the XCDM model, i.e., constant EoS for the dark energy component $w = constant$, performing the ESMCMC on the 7-dimensional parametric space $$\vec{\theta} \doteq (\Omega_c, H_0, w, \alpha, \beta, M_1, M_2).$$ Their best-fitting values and error bars are $\Omega_c = 0.269 \pm 0.017$, $H_0 = 71.04 \pm 1.59 \, \text{km} \, \text{s}^{-1} \text{Mpc}^{-1}$, $w = -1.11 \pm 0.07$, $\alpha = 0.142 \pm 0.007$, $\beta = 3.108 \pm 0.081$, $M_1 = -19.03 \pm 0.050$ and $M_2 = -19.10 \pm 0.05$. In figure~\ref{fig:real_qz} (right panel), we plotted the results, $\overline{q}^{\text{XCDM}} (z)$ (red line) and the CI's (red shaded areas), showing that XCDM model is also compatible within $99.73\%$ confidence level with the reconstructed curve in the entire redshift interval. It is worth noting that our conservative analyses provide only lower bounds for $z_{\tiny{T}} \simeq 0.19$ (JLA) and $\simeq 0.22$ [JLA+BAO+$H(z)$] within $99.73\%$. On the other hand, model-dependent works can estimate better constrained intervals for $z_{\tiny{T}}$ \cite{Capozziello2014, Santos2015}, as evinced by the XCDM results. However, as we discussed in sections~\ref{sec:introduction} and \ref{sec:review}, these estimates are accurate as long as the form-bias is small.

\begin{figure}[h]
\begin{center}
\includegraphics[scale=0.5]{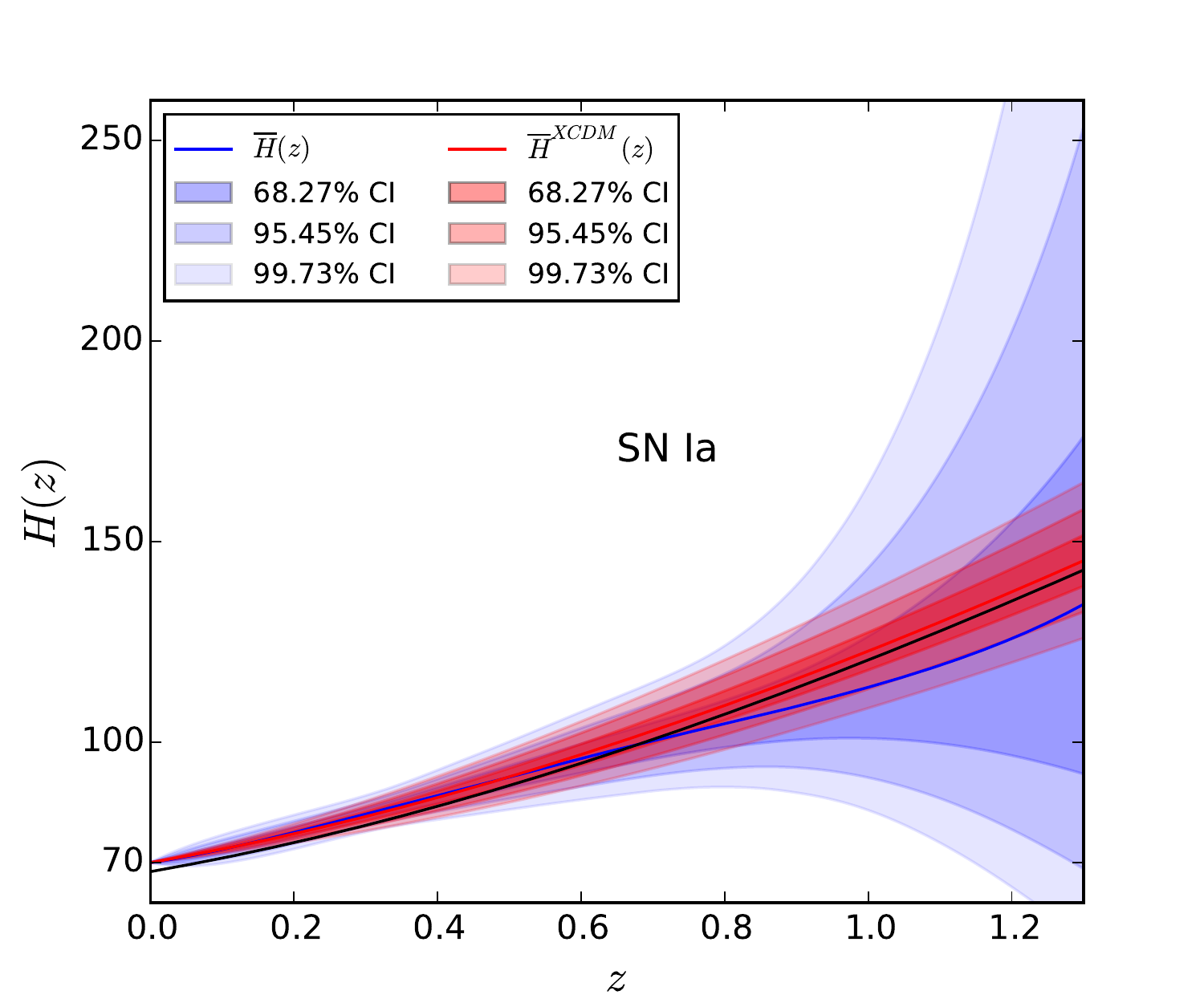}
\includegraphics[scale=0.5]{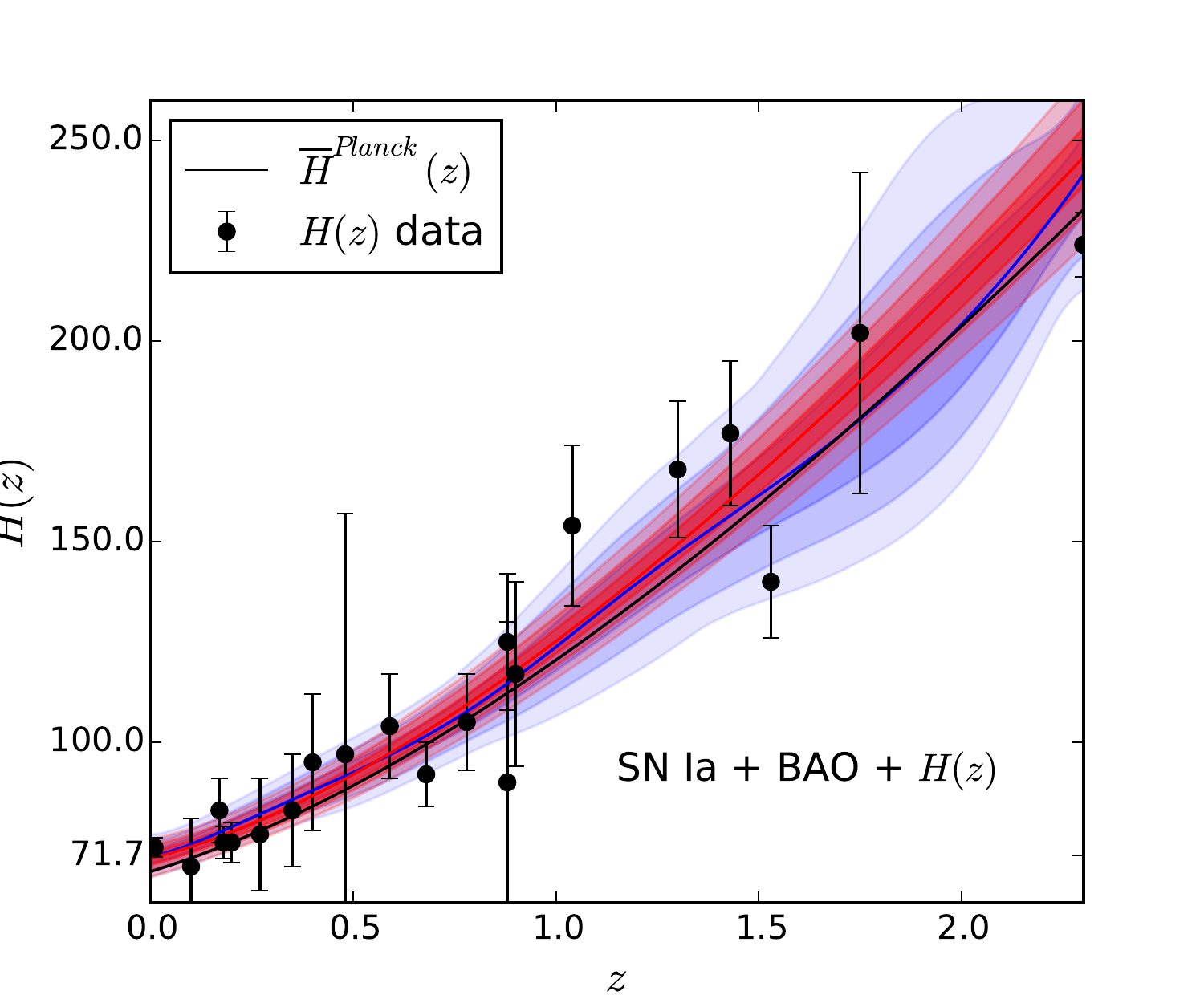}
\includegraphics[scale=0.5]{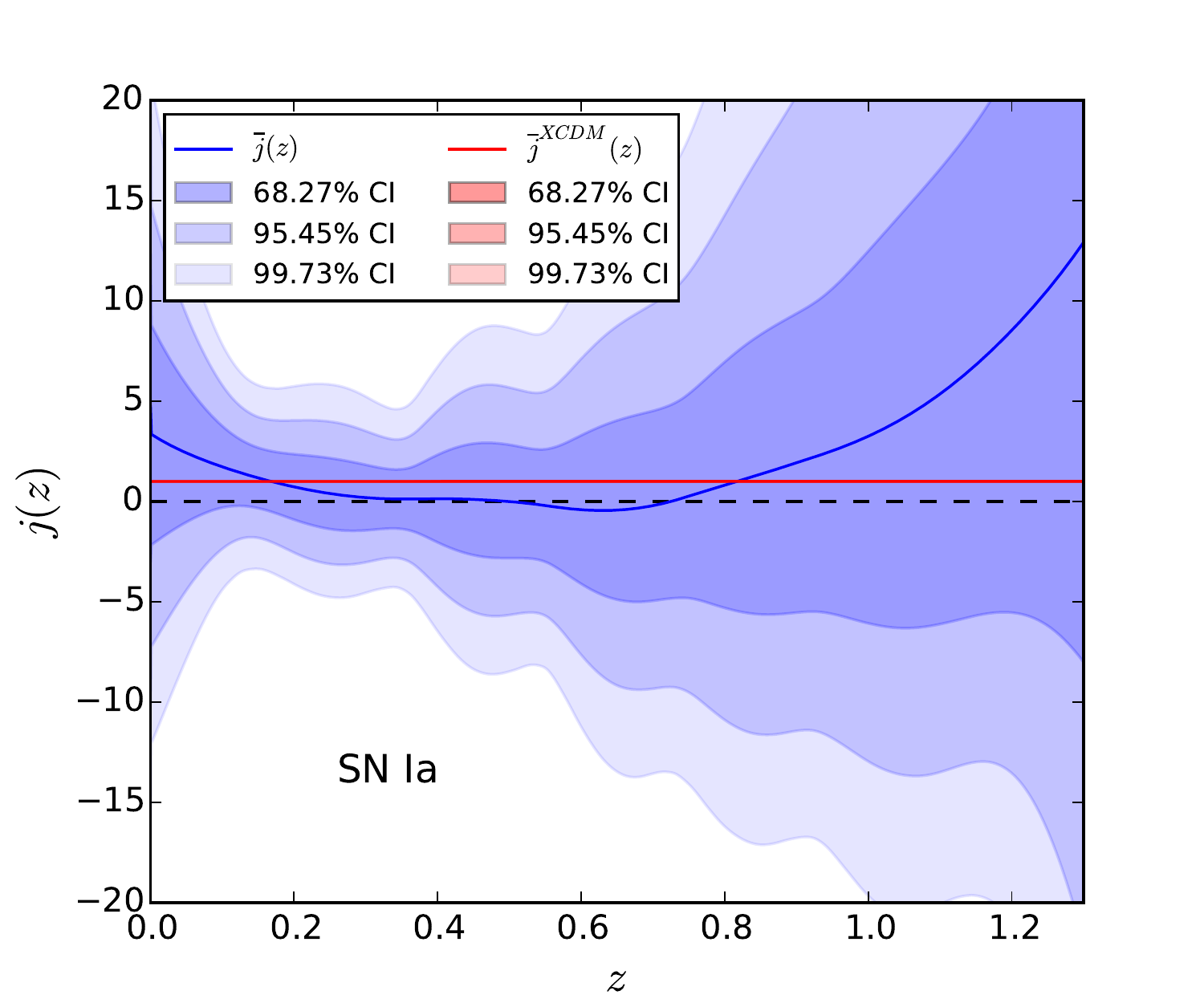}
\includegraphics[scale=0.5]{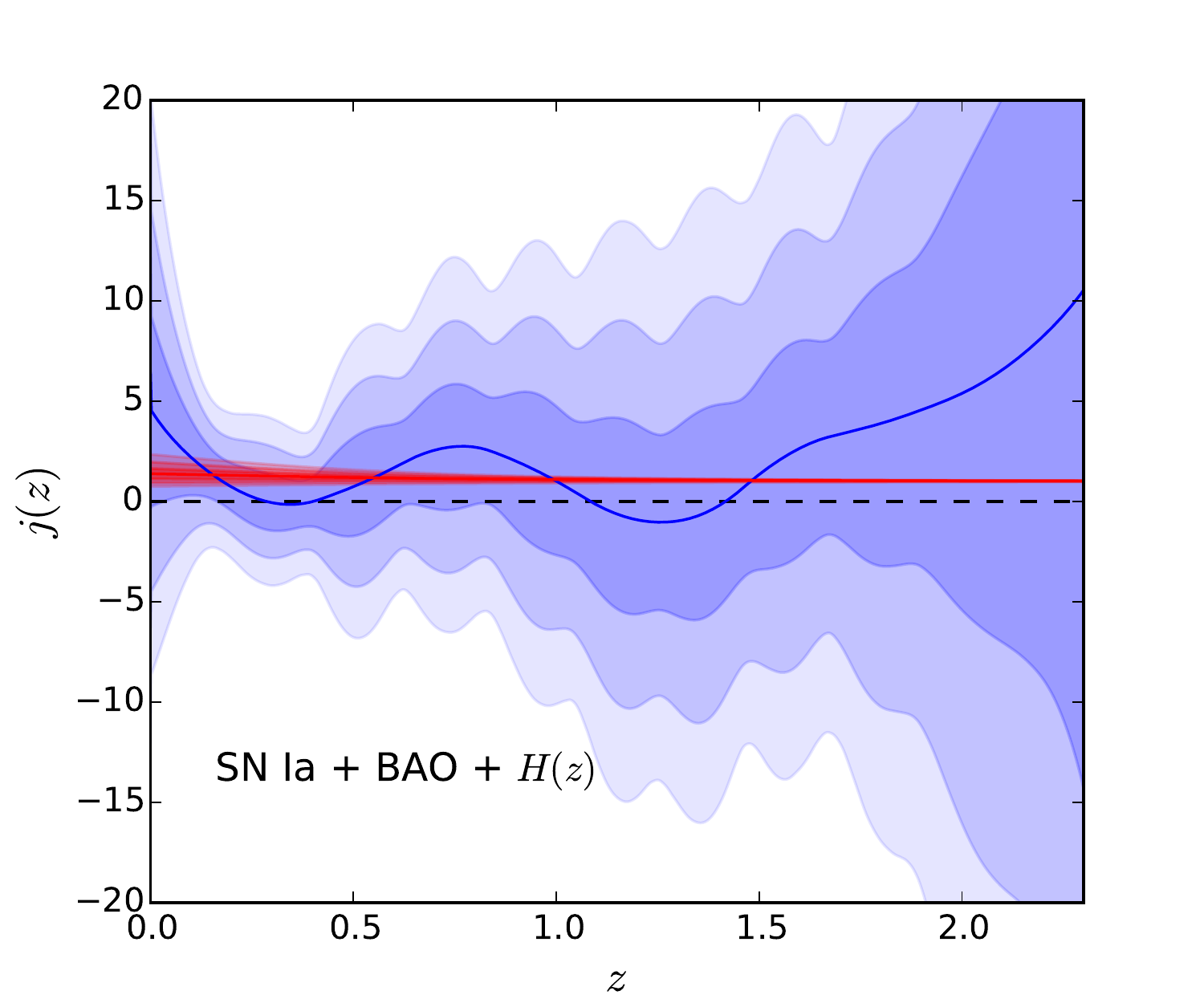}
\caption{$H(z)$ (upper panels) and $j(z)$ (lower panels) mean estimates (blue solid lines) and their $68.23\%$, $95.45\%$ and $99.73\%$ confidence intervals (blue shaded areas) obtained integrating and deriving, respectively, the reconstructed $q(z)$ using SNe Ia (left panels) and SNe Ia + BAO + $H(z)$ data (right panels). Similarly, the red line and contours are the $H(z)$ and $j(z)$ means and their CI's obtained integrating/deriving the respective $q(z)$ where the $\Lambda$CDM (left panels) and XCDM (right panels) models were assumed. The black lines correspond to the Planck+BAO+JLA+$H_0$ best-fit assuming $\Lambda$CDM model. The black dots and error bars represent the 21 $H(z)$ measurements used in this work.}
   \label{fig:real_Hz_jz}
\end{center}
\end{figure}

Besides the results presented so far, we can also infer other kinematic quantities from the reconstructed function $q(z)$ (see sections~\ref{sec:introduction} and \ref{sec:def_q}), such as $H(z)$ and the jerk function $j(z) \equiv - \dddot{a}/(aH^3)$ (i.e., the third order term of the scale factor expansion). The functions $H(z)$ and $j(z)$ are estimated by integrating and differentiating, respectively, $q(z)$ with respect to z. Figure~\ref{fig:real_Hz_jz} shows the results obtained using only JLA\footnote{In this case, we recover $H(z)$ [see eq.~\eqref{eq:Ez}] considering $H_0 = 70.0 \, \text{km}\,\text{s}^{-1}\text{Mpc}^{-1}$, which is the reference value used in \cite{Betoule2014}.} (left panels) and JLA+BAO+$H(z)$ (right panels). The upper panels display the means $\overline{H}(z)$ (blue lines) and the $68.23\%$, $95.45\%$ and $99.73\%$ CI's (blue shaded areas). 

As the $H(z)$ estimate is related with $q(z)$ by a numerical integration, it is consequently better constrained and the enhancement due to the combined data is even more apparent, as shown figure~\ref{fig:real_Hz_jz} (upper right panel), where the 21 $H(z)$ measurements are also plotted. As before, we compare these results with the respective outcomes assuming, respectively, $\Lambda$CDM and XCDM models (red lines and shaded areas). The decrease in the CI's of the reconstructed $H(z)$ highlights its concordance with those cosmological models within $99.73\%$ level. The Planck best-fit $\overline{H}^{\text{Planck}}(z)$ (black lines) also lies inside the CI's over the entire redshift interval for JLA+BAO+H(z), and it is outside the $99.73\% $ CI only in $0.0 \leq z \lesssim 0.03$ when we use JLA. 

Contrarily to the above scenario, $j(z)$ is obtained by numerical differentiation implying in high degenerated results, as displayed in the lower panels of figure~\ref{fig:real_Hz_jz}. The means $\overline{j}(z)$ and the CI's are represented by the blue lines and blue shaded areas, respectively. The estimated jerk functions, obtained assuming $\Lambda$CDM (left panel) and XCDM models (right panel), correspond to the red lines and red shaded areas. Besides the redshift intervals $[0.14, 0.39]$ (left panel) and $[0.13, 0.45]$ (right panel), which provides $-6 \lesssim j(z) \lesssim 6$ with $\geq 99.73\%$ significance, the degenerated results do not provide relevant information about this kinematic function.  
 
At last, in this work we also obtain a model-independent estimate of the sound horizon, since $r_d$ is fitted throughout the reconstruction procedure. Thus, we obtain $r_d = 101.15 \pm 1.8 \, h^{-1}\text{Mpc}$ given the combined analysis in the 18-dimensional parametric space using JLA+BAO+$H(z)$ data. Our result is in accordance with that presented in reference~\cite{Heavens2014}, where they obtained $r_d = 100.7 \pm 2.0 \, h^{-1}\text{Mpc}$ (SNe+BAO+$H(z)$ without Hubble prior) and $r_d = 101.9 \pm 1.9 \, h^{-1}\text{Mpc}$ (idem with Hubble prior) assuming a linear spline for $h^{-1}(z) = 100 \, \text{km}\,\text{s}^{-1}\text{Mpc}^{-1}/ H(z)$, 6 knots equally-spaced and curved universe. It is worth noting that their reconstruction was directly applied to the data and no test regarding the assumed curve was performed.


\section{Conclusions}
\label{sec:conclusions}

In this work we presented a general model-independent approach to reconstruct any one-variable function. In particular, we applied it to estimate the deceleration function $q(z)$ using the JLA SN Ia, BAO and $H(z)$ data sets. We performed a conservative analysis in the sense that we reconstructed $q(z)$ considering minimal assumptions (FLRW metric and flat universe) and also fitting simultaneously $H_0$, $r_d$ and the SN Ia astrophysical parameters ($\alpha$, $\beta$, $M_1$, $M_2$).

The MC results presented in sections~\ref{sec:mc_sneia} and \ref{sec:mc_alldata} show that it is crucial to estimate both the form- and the estimator-biases in order to validate any proposed reconstruction method, independently if it is parametric or non-parametric. They also reveal that the MSE's are independent of the underlying fiducial model and are dominated by the variance of the current data sets, even for cases where the reconstructed functions are far to be a good representation of the true curves, i.e., highly biased. Nevertheless, a blindly minimization of the MSE could lead to a bias dominated reconstruction. Such reconstructions must be carefully analyzed and the bias added by hand or estimated in some way. Even so, the information about the shape of the curve is mostly lost. Thus, requiring the bias to be at most $10\%$ of the total MSE, we evaluated the bias-variance trade-off obtaining that, currently, the best penalty factor is given by $\rel = 30\%$. 

Even though the MSE is approximately independent of the fiducial model, analyzing MSE of only one model can lead to misleading conclusions. If we had only used the $\Lambda$CDM fiducial model, the smaller $\rel = 5\%$ would prove to be the best choice, i.e., smaller MSE and insignificant bias. In other words, if a method is good in reconstructing a $\Lambda$CDM shaped curve, this does not mean that the same method will be able to reconstruct other curves close to it. In addition, capping the bias at $10\%$ is a straightforward way to obtain the total uncertainty of the estimates, since in this case we can use directly the variance analysis without requiring additional bias estimation.

Our main reconstructions are summarized in figure~\ref{fig:real_qz}, given that there are no assumptions on the gravitational dynamics nor on the matter content, and we also include the relevant astrophysical parameters into the study. The standard cosmological model agrees with the reconstructed results within $99.73\%$ CI in the entire redshift intervals. This is even further evident observing the estimate for the Hubble function in figure~\ref{fig:real_Hz_jz}. In this sense this work discards large deviations from the standard model with the caveat that we are assuming the FLRW metric and we are not testing this hypothesis. Notwithstanding, with more data available, we advocate that to measure departures from the standard model, one should repeat the process described in this work, evaluating the capability of the model to differentiate between a whole class of fiducial models. 


\acknowledgments 

SDPV thanks CAPES for the grant 2649-13-6. 
MPL thanks CNPq (PCI/MCTI/INPE program and grant 202131/2014-9) for financial support.
This research was performed using the Mesu-UV supercomputer of the Pierre \& Marie Curie University  -- France (UPMC) and the computer cluster of the State University of the Rio Grande do Norte (UERN) -- Brazil. We also thank Pierre Astier and Marc Betoule for kindly providing the complete covariance matrix necessary for the SNe Ia analysis, and Nicol\'{a}s Busca for useful comments.

\appendix 
\section{Simulated data}
\label{app:sampling}

In this appendix we describe the methodology to generate the SN Ia, BAO and $H(z)$ mock catalogs which we used 
to perform the MC study in section~\ref{sec:meth_valid}.

\subsection{SN Ia sampling}

The SN Ia distance modulus $\mu$ is written in terms of its light-curve parameters $(m_B, X, \mathcal{C})$ as 
\begin{equation}\label{eq:dist_mod}
\mu = m_B - (M_{h_i} - \alpha X + \beta \mathcal{C}),
\end{equation}
where $m_B$ is the observed peak magnitude in rest-frame B, $X$ and $\mathcal{C}$ are the observed stretching and the color at maximum brightness, respectively, and $\alpha$, $\beta$ and $M_{h_i} = \{M_1, M_2\}$ are nuisance parameters \cite{Betoule2014}. Thus, in order to generate SNe Ia samples, we must provide fiducial values of $(m_{B}, X, \mathcal{C})$, for each SN Ia, and their respective probability distributions. The fiducial values are usually determined choosing a specific set of the observable model parameters. In this case, from eq.~\eqref{eq:dist_mod} and the theoretical distance modulus, $\mu^{\text{th}} = 5\log_{10}(\mathcal{D}_L(z^{\text{hel}}, z^{{\text{cmb}}}) H_0/ c) + 25$, we can write the fiducial magnitude as function of $X_i$ and $\mathcal{C}_i$,
\begin{equation}\label{eq:mb_sneia}
m_{B_i}^\f = 5\log_{10}\left[\mathcal{D}_L\left(z^{\text{hel}}_i, z^{{\text{cmb}}}_i, \hat{q}^\f\right)\right] - \alpha^\f X_i + \beta^\f\mathcal{C}_i + M_{h_i}^\f - 5\log_{10}(c/H_0^\f) + 25,
\end{equation}
where $i$ denotes the SN Ia index and the SN Ia redshift values $\{z^{\text{hel}}_i, z^{{\text{cmb}}}_i\}$ are always kept fixed to their observable estimates. As described in section~\ref{sec:mc_sneia}, the fiducial parameters are 
$$\{\alpha^\f, \beta^\f, M_1^\f, M_2^\f, H_0^\f\} = \{0.141, 3.101, -19.05, -19.12, 73.0\},$$
and they correspond (besides $H_0^\f$) to the best-fitting values assuming $\Lambda$CDM model obtained in reference~\cite{Betoule2014}. We consider the three different $q^\f(z)$ represented in figure~\ref{fig:fiducial_models}. At this point, we still have to define the fiducial values for $X_i$ and $\mathcal{C}_i$.

However,  there is no astrophysical models for $X$ and $\mathcal{C}$. Therefore, we define the SNe Ia fiducial model $(m_{B_i}^\f, X_i^\f, \mathcal{C}_i^\f)$ as being the best-fitting values obtained by maximizing, with respect to all $X_i$ and $\mathcal{C}_i$, the multivariate Gaussian distribution~\cite{Betoule2014}
\begin{equation}\label{eq:sn_all_dist}
G(\vec{\zeta}, \mathsf{C}_{cmp}) = \frac{1}{\sqrt{(2\pi)^{3N} \vert\mathsf{C}_{cmp}\vert}} e^{-\frac{1}{2}\vec{\zeta}^T \mathsf{C}_{cmp}^{-1} \vec{\zeta}},
\end{equation}
where $\vert ... \vert$ denotes the determinant, $\mathsf{C}_{cmp}^{-1}$ is the complete inverse covariance matrix (with 2220 rows and columns), $N$ is the number of SNe Ia of the JLA sample (740) and 
$$\vec{\zeta} = \left(m_{B_1}, ..., m_{B_N}, X_{N+1}, ..., X_{2N}, \mathcal{C}_{2N+1}, ..., \mathcal{C}_{3N}\right),$$
with $\{m_{B_i}\}$ given by eq.~\eqref{eq:mb_sneia}.\footnote{The covariance matrix $\mathsf{C}_{cmp}$ also depends on the intrinsic standard deviation of each SN Ia subsample, i.e., SDSS, SNLS3, low-z and HST (Hubble Space Telescope), $\{\sigma^1_{int}, \sigma^2_{int}, \sigma^3_{int}, \sigma^4_{int}\}$, respectively. In particular, we used original values calibrated in~\cite{Betoule2014}.}

Finally, writing the SNe Ia variables as
\begin{equation}
m_{B_i} = m_{B_i}^\f + \delta m_{B_i}, \quad X_i = X^\f_i + \delta X_i \quad \text{and} \quad \mathcal{C}_i = \mathcal{C}^\f_i + \delta\mathcal{C}_i,
\end{equation}
we create a SN Ia sample $\{m_{B_i}, X_i, \mathcal{C}_i\}$ by randomly generating 2220 values from the normal distribution with variance $\mathsf{C}_{cmp}$ and zero mean corresponding to $$\overrightarrow{\delta\zeta} = \left(\delta m_{B_1}, ..., \delta m_{B_N}, \delta X_{N+1}, ..., \delta X_{2N}, \delta\mathcal{C}_{2N+1}, ..., \delta\mathcal{C}_{3N}\right).$$

\begin{table}[h]
\caption{BAO data} 
\label{tab:bao}
\centering 
\begin{threeparttable}[b]
\begin{tabular}{c c c c rrrrrr} 
\hline \hline 
Reference & z & $D_V(z)/r_d$ & \multicolumn{6}{c}{$\mathsf{C}_{\text{BAO}}^{-1}$} \\[0.5ex]
\hline
Beutler et al.\footnotemark[1] & 0.106 & $0.336$ & $0.015^{-2}$ & 0 & 0 & 0 & 0 & 0 \\ 
Padmanabhan et al. & 0.35 & 8.88 & 0 & 34.60 & 0 & 0 & 0 & 0  \\
\multirow{3}{*}{Kazin et al.\footnotemark[2]} & 0.44 & 11.550 & 0 & 0 & 4.8116 & -2.4651 & 1.0375 & 0 \\  
 & 0.60 & 14.945 & 0 &0 & -2.4651 & 3.7697 & -1.5865& 0 \\
 & 0.73 & 16.932 & 0 &0 & 1.0375 & -1.5865 & 3.6498 & 0 \\
Ross et al.\footnotemark[2] & 0.15 & 4.466 & - & - & - & - & - & - \\[1ex] 
\hline 
\end{tabular}
\begin{tablenotes}
    \item[1] In Ref.~\cite{Beutler2011} the authors provide $r_d / D_V(z)$. Note that it is the reciprocal of the provided in the other references. As this data is not correlated with the others, we can include it directly in $L_{BAO}$ [eq.~\eqref{eq:lik_bao}].
    \item[2] Refs.~\cite{Kazin2014} and \cite{Ross2014} provide $D_V(z)(r_d^{\f}/r_d)$, where $r_d^{\f} = 148.6$ and $148.69 \, h^{-1}\text{Mpc}$, respectively. We consider these values to build $L_{\text{BAO}}$.
  \end{tablenotes}
 \end{threeparttable}
\end{table}

\subsection{$H(z)$ and BAO sampling}

Differently from the SN Ia sampling, the BAO and $H(z)$ mock catalogs can be directly generated, since we have theoretical models for the respective observables, i.e., $D_V(z)/r_d$ [or $r_d/D_V(z)$], $H(z)$ and $H(z)r_d /(1+z)$, as displayed in eqs.~\eqref{eq:bao_dv}, \eqref{eq:bao_rd} and \eqref{eq:Ez}. 

Each $H^\text{obs}$ catalog correspond to $\{H_i^\text{obs}\}$, where $i = 1,..., 20$, and the point $\{H_r^{\text{obs}}\}$. The 21 redshift values are not altered in the sampling procedure and they are equal to the observed data $z$ of table~\ref{tab:Hz}. We obtain $H_i^\text{obs}$ writing it as $$H_i^\text{obs} = H^\f(z_i) + \delta H_i,$$ and, then,  we randomly generate the quantity $\delta H_i$ from a Gaussian distribution with mean 0 and standard deviation $\sigma_i$ (last column of table~\ref{tab:Hz}), since $H^\text{obs}_i$ follows a Gaussian distribution. Analogously, we generate the point $H^\text{obs}_r$ using a zero mean Gaussian with standard deviation $\sigma_{21}$ and add the result to $H(z_{21})r_d /(1+z_{21})$. The value of $H^\f(z_i)$ correspond to the fiducial value, which is defined by $q^\f(z)$, $H_0^\f = 73.0 \, \text{km} \, \text{s}^{-1} \text{Mpc}^{-1}$ and $r_d^\f = 103.5 \, h^{-1}\text{Mpc}$. 

Applying the same methodology, we generate the first two points of the BAO set $\{D_{V_j}/r_d\}$, where $D_V(z_j)/r_d = D_V^\f(z_j)/r_d^\f + \delta{}b_j$, the fiducial values are also defined by $q^\f(z)$, $H_0^\f$ and $r_d^\f$, and their respective inverse variances are presented in the first two rows of table~\ref{tab:bao}. The three correlated BAO points, related to Kazin et al.~\cite{Kazin2014} data, are resampled randomly generating the 3 values $\delta{}b_j$ from a multivariate Gaussian distribution with means 0 and covariance matrix provided in reference~\cite{Kazin2014}, see table~\ref{tab:bao}. 

The last BAO point ($j = 6$) follows a different probability distribution, which is defined by the likelihood function $L_{Ross}$ provided by Ross et al.~\cite{Ross2014} and centered in the observed value $(D_{V_j}/r_d)^{\text{obs}} = 4.466$. Thus, to produce a mock catalog from a fiducial model we perform the inverse transform sampling. First, we shift this function, $L_{Ross}^s$ such that the new center is $D_V^\f(z_j)/r_d^\f$. Then, we randomly generate a number $u$ from a uniform distribution $(0, 1)$ and, finally, we invert the equation
\begin{equation}
u = \int_{\alpha_i}^\alpha \dd\alpha^\prime L_{Ross}^s (\alpha^\prime)
\end{equation}
obtaining $\alpha(u)$. The lower bound $\alpha_i$ corresponds to the lower bound described by the $L_{Ross}$ function.

\begin{table}[h]
\caption{$H(z)$ data} 
\label{tab:Hz}
\centering 
\begin{threeparttable}[b]
\begin{tabular}{c c c c | c c c c} 
\hline \hline 
Reference & z & $H(z)$ & $\sigma$ & Reference & z & $H(z)$ & $\sigma$ \\[0.5ex]
\hline
Riess et al.~\cite{Riess2011} & 0.0 & $73.8$ & 2.4 & \multirow{11}{*}{Stern et al.~\cite{Stern2010}} & 0.1 & 69 & 12 \\  
\multirow{8}{*}{Moresco et al.~\cite{Moresco2012}} & 0.18 & 75 & 4 & & 0.17 & 83 & 8 \\
& 0.20 & 75 & 5 & & 0.27 & 77 & 14 \\
& 0.35 & 83 & 14 & & 0.4 & 95 & 17 \\
& 0.59 & 104 & 13 & & 0.48 & 97 & 60 \\
& 0.68 & 92 & 8 & & 0.88 & 90 & 40 \\
& 0.78 & 105 & 12 & & 0.9 & 117 & 23 \\
& 0.88 & 125 & 17 & & 1.3 & 168 & 17 \\
& 1.04 & 154 & 20 & & 1.43 & 177 & 18 \\
& & &  & & 1.53 & 140 & 14 \\
Busca et al.~\cite{Busca2013}\footnotemark[1] & - & - & - & & 1.75 & 202 & 40 \\ 
\hline \hline
\end{tabular}
\begin{tablenotes}
    \item[1] The observable of this entry is related to $H(z_{21})r_d/(1+z_{21})$, instead of simply $H(z)$ as the other data points. In this entry the redshift is $z_{21} = 2.3$, the observable value and its standard deviation are $H^{\text{obs}}_r = 224 \, \text{km} \, \text{s}^{-1} \text{Mpc}^{-1}$ and $\sigma_{21} = 8 \, \text{km} \, \text{s}^{-1} \text{Mpc}^{-1}$, respectively.
  \end{tablenotes}
\end{threeparttable}
\end{table}


\begin{thebibliography}{67}

\bibitem{Riess1998}
{\bf Supernova Search Team} Collaboration, A.~G. Riess, A.~V. Filippenko,
  P.~Challis, A.~Clocchiattia, A.~Diercks, et~al., {\it Observational evidence
  from supernovae for an accelerating universe and a cosmological constant},
  {\em Astron. J.} {\bf 116} (1998) 1009--1038,
  [\href{http://xxx.lanl.gov/abs/astro-ph/9805201}{{\tt astro-ph/9805201}}].

\bibitem{Perlmutter1999}
{\bf Supernova Cosmology Project} Collaboration, S.~Perlmutter, G.~Aldering,
  G.~Goldhaber, R.~A. Knop, P.~Nugent, et~al., {\it Measurements of omega and
  lambda from 42 high-redshift supernovae},  {\em Astrophys. J.} {\bf 517}
  (1999) 565--586, [\href{http://xxx.lanl.gov/abs/astro-ph/9812133}{{\tt
  astro-ph/9812133}}].

\bibitem{Joyce2015}
A.~{Joyce}, B.~{Jain}, J.~{Khoury}, and M.~{Trodden}, {\it Beyond the
  cosmological standard model},  {\em Phys. Rep.} {\bf 568} (2015) 1--98,
  [\href{http://xxx.lanl.gov/abs/1407.0059}{{\tt arXiv:1407.0059}}].

\bibitem{Dvali2000}
G.~Dvali, G.~Gabadadze, and M.~Porrati, {\it 4d gravity on a brane in 5d
  minkowski space},  {\em Phys. Lett. B} {\bf 485} (2000) 208--214,
  [\href{http://xxx.lanl.gov/abs/hep-th/0005016}{{\tt hep-th/0005016}}].

\bibitem{Sotiriou2010}
T.~P. {Sotiriou} and V.~{Faraoni}, {\it f(r) theories of gravity},  {\em Rev.
  Mod. Phys.} {\bf 82} (2010) 451--497,
  [\href{http://xxx.lanl.gov/abs/0805.1726}{{\tt arXiv:0805.1726}}].

\bibitem{Nojiri2011}
S.~{Nojiri} and S.~D. {Odintsov}, {\it Unified cosmic history in modified
  gravity: From f(r) theory to lorentz non-invariant models},  {\em Phys. Rep.}
  {\bf 505} (2011) 59--144, [\href{http://xxx.lanl.gov/abs/1011.0544}{{\tt
  arXiv:1011.0544}}].

\bibitem{Turner2002c}
M.~S. Turner and A.~Riess, {\it Do sne ia provide direct evidence for past
  deceleration of the universe?},  {\em Astrophys. J.} {\bf 569} (2002) 18,
  [\href{http://xxx.lanl.gov/abs/astro-ph/0106051}{{\tt astro-ph/0106051}}].

\bibitem{Visser2004}
M.~Visser, {\it Jerk, snap, and the cosmological equation of state},  {\em
  Classical Quant. Grav.} {\bf 21} (2004) 2603--2616,
  [\href{http://xxx.lanl.gov/abs/gr-qc/0309109}{{\tt gr-qc/0309109}}].

\bibitem{Rapetti2007}
D.~Rapetti, S.~W. Allen, M.~A. Amin, and R.~D. Blandford, {\it A kinematical
  approach to dark energy studies},  {\em Mon. Not. R. Astron. Soc.} {\bf 375}
  (2007) 1510--1520, [\href{http://xxx.lanl.gov/abs/astro-ph/0605683}{{\tt
  astro-ph/0605683}}].

\bibitem{Zhai2013}
Z.-X. {Zhai}, M.-J. {Zhang}, Z.-S. {Zhang}, X.-M. {Liu}, and T.-J. {Zhang},
  {\it {Reconstruction and constraining of the jerk parameter from OHD and SNe
  Ia observations}},  {\em Phys. Lett. B} {\bf 727} (2013) 8--20,
  [\href{http://xxx.lanl.gov/abs/1303.1620}{{\tt arXiv:1303.1620}}].

\bibitem{Daly2003}
R.~A. {Daly} and S.~G. {Djorgovski}, {\it A model-independent determination of
  the expansion and acceleration rates of the universe as a function of
  redshift and constraints on dark energy},  {\em The Astrophys. J.} {\bf 597}
  (2003) 9--20, [\href{http://xxx.lanl.gov/abs/astro-ph/0305197}{{\tt
  astro-ph/0305197}}].

\bibitem{Benitez-Herrera2012}
S.~{Benitez-Herrera}, F.~{R{\"o}pke}, W.~{Hillebrandt}, C.~{Mignone},
  M.~{Bartelmann}, et~al., {\it Model-independent reconstruction of the
  expansion history of the universe from type ia supernovae},  {\em Mon. Not.
  R. Astron. Soc.} {\bf 419} (2012) 513--521,
  [\href{http://xxx.lanl.gov/abs/1109.0873}{{\tt arXiv:1109.0873}}].

\bibitem{Huterer2003}
D.~Huterer and G.~Starkman, {\it Parameterization of dark-energy properties: a
  principal-component approach},  {\em Phys. Rev. Lett.} {\bf 90} (2003)
  031301, [\href{http://xxx.lanl.gov/abs/astro-ph/0207517}{{\tt
  astro-ph/0207517}}].

\bibitem{Daly2008}
R.~A. {Daly}, S.~G. {Djorgovski}, K.~A. {Freeman}, M.~P. {Mory}, C.~P. {O'Dea},
  et~al., {\it Improved constraints on the acceleration history of the universe
  and the properties of the dark energy},  {\em Astrophys. J.} {\bf 677} (2008)
  1--11, [\href{http://xxx.lanl.gov/abs/0710.5345}{{\tt arXiv:0710.5345}}].

\bibitem{Sahni2006}
V.~{Sahni} and A.~{Starobinsky}, {\it Reconstructing dark energy},  {\em Int.
  J. Mod. Phys. D} {\bf 15} (2006) 2105--2132,
  [\href{http://xxx.lanl.gov/abs/astro-ph/0610026}{{\tt astro-ph/0610026}}].

\bibitem{Riess2004}
{\bf Supernova Search Team} Collaboration, A.~G. Riess, L.-G. Strolger,
  J.~Tonry, S.~Casertano, H.~C. Ferguson, et~al., {\it Type ia supernova
  discoveries at z>1 from the hubble space telescope: Evidence for past
  deceleration and constraints on dark energy evolution},  {\em Astrophys. J.}
  {\bf 607} (2004) 665--687,
  [\href{http://xxx.lanl.gov/abs/astro-ph/0402512}{{\tt astro-ph/0402512}}].

\bibitem{Shapiro2006}
C.~Shapiro and M.~S. Turner, {\it What do we really know about cosmic
  acceleration?},  {\em Astrophys. J.} {\bf 649} (2006) 563--569,
  [\href{http://xxx.lanl.gov/abs/astro-ph/0512586}{{\tt astro-ph/0512586}}].

\bibitem{Avgoustidis2009}
A.~{Avgoustidis}, L.~{Verde}, and R.~{Jimenez}, {\it Consistency among distance
  measurements: transparency, bao scale and accelerated expansion},  {\em J.
  Cosmol. Astropart. Phys.} {\bf 6} (2009) 12,
  [\href{http://xxx.lanl.gov/abs/0902.2006}{{\tt arXiv:0902.2006}}].

\bibitem{Nair2012}
R.~{Nair}, S.~{Jhingan}, and D.~{Jain}, {\it Cosmokinetics: a joint analysis of
  standard candles, rulers and cosmic clocks},  {\em J. Cosmol. Astropart.
  Phys.} {\bf 1} (2012) 18, [\href{http://xxx.lanl.gov/abs/1109.4574}{{\tt
  arXiv:1109.4574}}].

\bibitem{Clarkson2010}
C.~{Clarkson} and C.~{Zunckel}, {\it Direct reconstruction of dark energy},
  {\em Phys. Rev. Lett.} {\bf 104} (2010) 211301,
  [\href{http://xxx.lanl.gov/abs/1002.5004}{{\tt arXiv:1002.5004}}].

\bibitem{Ishida2011}
E.~E.~O. {Ishida} and R.~S. {de Souza}, {\it Hubble parameter reconstruction
  from a principal component analysis: minimizing the bias},  {\em Astron.
  Astrophys.} {\bf 527} (2011) A49,
  [\href{http://xxx.lanl.gov/abs/1012.5335}{{\tt arXiv:1012.5335}}].

\bibitem{Ruiz2012}
E.~J. {Ruiz}, D.~L. {Shafer}, D.~{Huterer}, and A.~{Conley}, {\it Principal
  components of dark energy with supernova legacy survey supernovae: The
  effects of systematic errors},  {\em Phys. Rev. D} {\bf 86} (2012) 103004,
  [\href{http://xxx.lanl.gov/abs/1207.4781}{{\tt arXiv:1207.4781}}].

\bibitem{Benitez-Herrera2013}
S.~{Benitez-Herrera}, E.~E.~O. {Ishida}, M.~{Maturi}, W.~{Hillebrandt},
  M.~{Bartelmann}, et~al., {\it {Cosmological parameter estimation from SN Ia
  data: a model-independent approach}},  {\em Mon. Not. R. Astron. Soc.} {\bf
  436} (2013) 854--858, [\href{http://xxx.lanl.gov/abs/1308.5653}{{\tt
  arXiv:1308.5653}}].

\bibitem{Shafieloo2006}
A.~{Shafieloo}, U.~{Alam}, V.~{Sahni}, and A.~A. {Starobinsky}, {\it Smoothing
  supernova data to reconstruct the expansion history of the universe and its
  age},  {\em Mon. Not. R. Astron. Soc.} {\bf 366} (2006) 1081--1095,
  [\href{http://xxx.lanl.gov/abs/astro-ph/0505329}{{\tt astro-ph/0505329}}].

\bibitem{Shafieloo2012}
A.~{Shafieloo}, {\it Crossing statistic: reconstructing the expansion history
  of the universe},  {\em J. Cosmol. Astropart. Phys.} {\bf 8} (2012) 2,
  [\href{http://xxx.lanl.gov/abs/1204.1109}{{\tt arXiv:1204.1109}}].

\bibitem{Holsclaw2010}
T.~{Holsclaw}, U.~{Alam}, B.~{Sans{\'o}}, H.~{Lee}, K.~{Heitmann}, et~al., {\it
  Nonparametric dark energy reconstruction from supernova data},  {\em Phys.
  Rev. Lett.} {\bf 105} (2010) 241302,
  [\href{http://xxx.lanl.gov/abs/1011.3079}{{\tt arXiv:1011.3079}}].

\bibitem{Seikel2012}
M.~Seikel, C.~Clarkson, and M.~Smith, {\it Reconstruction of dark energy and
  expansion dynamics using gaussian processes},  {\em J. Cosmol. Astropart.
  Phys.} {\bf 6} (2012) 36, [\href{http://xxx.lanl.gov/abs/1204.2832}{{\tt
  arXiv:1204.2832}}].

\bibitem{Montiel2014}
A.~{Montiel}, R.~{Lazkoz}, I.~{Sendra}, C.~{Escamilla-Rivera}, and
  V.~{Salzano}, {\it Nonparametric reconstruction of the cosmic expansion with
  local regression smoothing and simulation extrapolation},  {\em Phys. Rev. D}
  {\bf 89} (2014) 043007, [\href{http://xxx.lanl.gov/abs/1401.4188}{{\tt
  arXiv:1401.4188}}].

\bibitem{Conley2011}
A.~Conley, J.~Guy, M.~Sullivan, N.~Regnault, P.~Astier, et~al., {\it Supernova
  constraints and systematic uncertainties from the first three years of the
  supernova legacy survey},  {\em Astrophys. J. Suppl. Ser.} {\bf 192} (2011)
  1, [\href{http://xxx.lanl.gov/abs/1104.1443}{{\tt arXiv:1104.1443}}].

\bibitem{Betoule2014}
M.~{Betoule}, R.~{Kessler}, J.~{Guy}, J.~{Mosher}, D.~{Hardin}, et~al., {\it
  Improved cosmological constraints from a joint analysis of the sdss-ii and
  snls supernova samples},  {\em Astron. Astrophys.} {\bf 568} (2014) A22,
  [\href{http://xxx.lanl.gov/abs/1401.4064}{{\tt arXiv:1401.4064}}].

\bibitem{Beutler2011}
F.~{Beutler}, C.~{Blake}, M.~{Colless}, D.~H. {Jones}, L.~{Staveley-Smith},
  et~al., {\it The 6df galaxy survey: baryon acoustic oscillations and the
  local hubble constant},  {\em Mon. Not. R. Astron. Soc.} {\bf 416} (2011)
  3017--3032, [\href{http://xxx.lanl.gov/abs/1106.3366}{{\tt
  arXiv:1106.3366}}].

\bibitem{Padmanabhan2012}
N.~{Padmanabhan}, X.~{Xu}, D.~J. {Eisenstein}, R.~{Scalzo}, A.~J. {Cuesta},
  et~al., {\it A 2 per cent distance to z = 0.35 by reconstructing baryon
  acoustic oscillations - i. methods and application to the sloan digital sky
  survey},  {\em Mon. Not. R. Astron. Soc.} {\bf 427} (2012) 2132--2145,
  [\href{http://xxx.lanl.gov/abs/1202.0090}{{\tt arXiv:1202.0090}}].

\bibitem{Kazin2014}
E.~A. {Kazin}, J.~{Koda}, C.~{Blake}, N.~{Padmanabhan}, S.~{Brough}, et~al.,
  {\it The wigglez dark energy survey: improved distance measurements to z = 1
  with reconstruction of the baryonic acoustic feature},  {\em Mon. Not. R.
  Astron. Soc.} {\bf 441} (2014) 3524--3542,
  [\href{http://xxx.lanl.gov/abs/1401.0358}{{\tt arXiv:1401.0358}}].

\bibitem{Ross2014}
A.~J. {Ross}, L.~{Samushia}, C.~{Howlett}, W.~J. {Percival}, A.~{Burden},
  et~al., {\it The clustering of the sdss dr7 main galaxy sample i: A 4 per
  cent distance measure at z=0.15},
  \href{http://xxx.lanl.gov/abs/1409.3242}{{\tt arXiv:1409.3242}}.

\bibitem{Stern2010}
D.~{Stern}, R.~{Jimenez}, L.~{Verde}, M.~{Kamionkowski}, and S.~A. {Stanford},
  {\it Cosmic chronometers: constraining the equation of state of dark energy.
  i: H(z) measurements},  {\em J. Cosmol. Astropart. Phys.} {\bf 2} (2010) 8,
  [\href{http://xxx.lanl.gov/abs/0907.3149}{{\tt arXiv:0907.3149}}].

\bibitem{Riess2011}
A.~G. {Riess}, L.~{Macri}, S.~{Casertano}, H.~{Lampeitl}, H.~C. {Ferguson},
  et~al., {\it A 3
  hubble space telescope and wide field camera 3},  {\em Astrophys. J.} {\bf
  730} (2011) 119, [\href{http://xxx.lanl.gov/abs/1103.2976}{{\tt
  arXiv:1103.2976}}].

\bibitem{Moresco2012}
M.~{Moresco}, A.~{Cimatti}, R.~{Jimenez}, L.~{Pozzetti}, G.~{Zamorani}, et~al.,
  {\it Improved constraints on the expansion rate of the universe up to z \~{}
  1.1 from the spectroscopic evolution of cosmic chronometers},  {\em J.
  Cosmol. Astropart. Phys.} {\bf 8} (2012) 6,
  [\href{http://xxx.lanl.gov/abs/1201.3609}{{\tt arXiv:1201.3609}}].

\bibitem{Busca2013}
N.~G. Busca, T.~Delubac, J.~Rich, S.~Bailey, A.~Font-Ribera, et~al., {\it
  Baryon acoustic oscillations in the ly{$\alpha$} forest of boss quasars},
  {\em Astron. Astrophys.} {\bf 552} (2013) A96,
  [\href{http://xxx.lanl.gov/abs/1211.2616}{{\tt arXiv:1211.2616}}].

\bibitem{Hinshaw2013}
G.~{Hinshaw}, D.~{Larson}, E.~{Komatsu}, D.~N. {Spergel}, C.~L. {Bennett},
  et~al., {\it Nine-year wilkinson microwave anisotropy probe (wmap)
  observations: Cosmological parameter results},  {\em Astrophys. J. Suppl.
  Ser.} {\bf 208} (2013) 19, [\href{http://xxx.lanl.gov/abs/1212.5226}{{\tt
  arXiv:1212.5226}}].

\bibitem{Planck2015}
{Planck Collaboration}, P.~A.~R. {Ade}, N.~{Aghanim}, M.~{Arnaud},
  M.~{Ashdown}, et~al., {\it Planck 2015 results. xiii. cosmological
  parameters},  \href{http://xxx.lanl.gov/abs/1502.0158}{{\tt
  arXiv:1502.0158}}.

\bibitem{Komatsu2011}
E.~Komatsu, K.~M. Smith, J.~Dunkley, C.~L. Bennett, B.~Gold, et~al., {\it
  Seven-year wilkinson microwave anisotropy probe (wmap) observations:
  Cosmological interpretation},  {\em Astrophys. J. Suppl. Ser.} {\bf 192}
  (2011) 18, [\href{http://xxx.lanl.gov/abs/1001.4538}{{\tt arXiv:1001.4538}}].

\bibitem{Lima2008}
M.~P. Lima, S.~Vitenti, and M.~J. Rebouças, {\it Energy conditions bounds and
  their confrontation with supernovae data},  {\em Phys. Rev. D} {\bf 77}
  (2008) 083518, [\href{http://xxx.lanl.gov/abs/0802.0706}{{\tt
  arXiv:0802.0706}}].

\bibitem{Lima2008a}
M.~P. Lima, S.~D.~P. Vitenti, and M.~J. Rebouças, {\it Energy conditions
  bounds and supernovae data},  {\em Phys. Lett. B} {\bf 668} (2008) 83,
  [\href{http://xxx.lanl.gov/abs/0808.2467}{{\tt arXiv:0808.2467}}].

\bibitem{Cattoeen2008}
C.~{Catto{\"e}n} and M.~{Visser}, {\it Cosmographic hubble fits to the
  supernova data},  {\em Phys. Rev. D} {\bf 78} (2008) 063501,
  [\href{http://xxx.lanl.gov/abs/0809.0537}{{\tt arXiv:0809.0537}}].

\bibitem{Penna-Lima2014}
M.~{Penna-Lima}, M.~{Makler}, and C.~A. {Wuensche}, {\it Biases on cosmological
  parameter estimators from galaxy cluster number counts},  {\em J. Cosmol.
  Astropart. Phys.} {\bf 5} (2014) 39,
  [\href{http://xxx.lanl.gov/abs/1312.4430}{{\tt arXiv:1312.4430}}].

\bibitem{Crittenden2012}
R.~G. {Crittenden}, G.-B. {Zhao}, L.~{Pogosian}, L.~{Samushia}, and X.~{Zhang},
  {\it {Fables of reconstruction: controlling bias in the dark energy equation
  of state}},  {\em J. Cosmol. Astropart. Phys.} {\bf 2} (2012) 48,
  [\href{http://xxx.lanl.gov/abs/1112.1693}{{\tt arXiv:1112.1693}}].

\bibitem{Zhao2012}
G.-B. {Zhao}, R.~G. {Crittenden}, L.~{Pogosian}, and X.~{Zhang}, {\it
  {Examining the Evidence for Dynamical Dark Energy}},  {\em Phys. Rev. Lett.}
  {\bf 109} (2012) 171301, [\href{http://xxx.lanl.gov/abs/1207.3804}{{\tt
  arXiv:1207.3804}}].

\bibitem{Daly2004}
R.~A. Daly and S.~G. Djorgovski, {\it Direct determination of the kinematics of
  the universe and properties of the dark energy as functions of redshift},
  {\em Astrophys. J.} {\bf 612} (2004) 652--659,
  [\href{http://xxx.lanl.gov/abs/astro-ph/0403664}{{\tt astro-ph/0403664}}].

\bibitem{Boor2001}
C.~de~Boor, {\em A Practical Guide to Splines}.
\newblock Springer, 2001.

\bibitem{Bishop1996}
C.~M. Bishop, {\em Neural Networks for Pattern Recognition}.
\newblock Oxford University Press, 1996.

\bibitem{Keijzer2000}
M.~Keijzer and V.~Babovic, {\it Genetic programming, ensemble methods and the
  bias/variance tradeoff – introductory investigations},  in {\em Genetic
  Programming} (R.~Poli, W.~Banzhaf, W.~Langdon, J.~Miller, P.~Nordin, et~al.,
  eds.), vol.~1802 of {\em Lecture Notes in Computer Science}, pp.~76--90.
\newblock Springer Berlin Heidelberg, 2000.

\bibitem{Wasserman2001}
L.~Wasserman, C.~J. Miller, R.~C. Nichol, C.~Genovese, W.~Jang, et~al., {\it
  Non-parametric inference in astrophysics},
  \href{http://xxx.lanl.gov/abs/astro-ph/0112050}{{\tt astro-ph/0112050}}.

\bibitem{DiasPintoVitenti2014}
S.~{Dias Pinto Vitenti} and M.~{Penna-Lima}, {\it Numcosmo: Numerical
  cosmology},  {\em Astrophysics Source Code Library} (2014)
  [\href{http://xxx.lanl.gov/abs/ascl:1408.013}{{\tt ascl:1408.013}}].

\bibitem{Davis2011}
T.~M. {Davis}, L.~{Hui}, J.~A. {Frieman}, T.~{Haugb{\o}lle}, R.~{Kessler},
  et~al., {\it The effect of peculiar velocities on supernova cosmology},  {\em
  Astrophys. J.} {\bf 741} (2011) 67,
  [\href{http://xxx.lanl.gov/abs/1012.2912}{{\tt arXiv:1012.2912}}].

\bibitem{Thepsuriya2014}
K.~{Thepsuriya} and A.~{Lewis}, {\it Accuracy of cosmological parameters using
  the baryon acoustic scale},  \href{http://xxx.lanl.gov/abs/1409.5066}{{\tt
  arXiv:1409.5066}}.

\bibitem{Aubourg2014}
{\'E}.~{Aubourg}, S.~{Bailey}, J.~E. {Bautista}, F.~{Beutler}, V.~{Bhardwaj},
  et~al., {\it Cosmological implications of baryon acoustic oscillation (bao)
  measurements},  \href{http://xxx.lanl.gov/abs/1411.1074}{{\tt
  arXiv:1411.1074}}.

\bibitem{Eisenstein2005}
{\bf SDSS} Collaboration, D.~J. Eisenstein, I.~Zehavi, D.~W. Hogg,
  R.~Scoccimarro, M.~R. Blanton, et~al., {\it Detection of the baryon acoustic
  peak in the large-scale correlation function of sdss luminous red galaxies},
  {\em Astrophys. J.} {\bf 633} (2005) 560--574,
  [\href{http://xxx.lanl.gov/abs/astro-ph/0501171}{{\tt astro-ph/0501171}}].

\bibitem{Eisenstein2007}
D.~J. {Eisenstein}, H.-J. {Seo}, E.~{Sirko}, and D.~N. {Spergel}, {\it
  Improving cosmological distance measurements by reconstruction of the baryon
  acoustic peak},  {\em Astrophys. J.} {\bf 664} (2007) 675--679,
  [\href{http://xxx.lanl.gov/abs/astro-ph/0604362}{{\tt astro-ph/0604362}}].

\bibitem{Eisenstein1998}
D.~J. Eisenstein and W.~Hu, {\it Baryonic features in the matter transfer
  function},  {\em Astrophys. J.} {\bf 496} (1998) 605,
  [\href{http://xxx.lanl.gov/abs/astro-ph/9709112}{{\tt astro-ph/9709112}}].

\bibitem{Lewis2000}
A.~Lewis, A.~Challinor, and A.~Lasenby, {\it Efficient computation of {CMB}
  anisotropies in closed {FRW} models},  {\em Astrophys. J.} {\bf 538} (2000)
  473--476, [\href{http://xxx.lanl.gov/abs/astro-ph/9911177}{{\tt
  astro-ph/9911177}}].

\bibitem{Efron1994}
B.~Efron and R.~Tibshirani, {\em An Introduction to the Bootstrap}.
\newblock Chapman \& Hall/CRC Monographs on Statistics \& Applied Probability.
  Taylor \& Francis, 1994.

\bibitem{Goodman2010}
J.~Goodman and J.~Weare, {\it Ensemble samplers with affine invariance},  {\em
  CAMCoS} {\bf 5} (2010) 65--80.

\bibitem{Foreman-Mackey2013}
D.~Foreman-Mackey, D.~W. Hogg, D.~Lang, and J.~Goodman, {\it emcee : The {MCMC}
  hammer},  {\em Publ. Astron. Soc. Pac.} {\bf 125} (2013) 306--312.

\bibitem{Brooks1998}
S.~P. Brooks and A.~Gelman, {\it General methods for monitoring convergence of
  iterative simulations},  {\em J. Comput. Graph. Stat.} {\bf 7} (1998)
  434--455.

\bibitem{Capozziello2014}
S.~{Capozziello}, O.~{Farooq}, O.~{Luongo}, and B.~{Ratra}, {\it {Cosmographic
  bounds on the cosmological deceleration-acceleration transition redshift in
  f(R) gravity}},  {\em Phys. Rev. D} {\bf 90} (2014) 044016,
  [\href{http://xxx.lanl.gov/abs/1403.1421}{{\tt arXiv:1403.1421}}].

\bibitem{Santos2015}
M.~V. dos Santos, R.~R.~R. Reis, and I.~Waga, {\it Constraining cosmic
  deceleration-acceleration transition with type ia supernova, bao/cmb and h(z)
  data},  \href{http://xxx.lanl.gov/abs/1505.0381}{{\tt arXiv:1505.0381}}.

\bibitem{Heavens2014}
A.~{Heavens}, R.~{Jimenez}, and L.~{Verde}, {\it Standard rulers, candles, and
  clocks from the low-redshift universe},  {\em Phys. Rev. Lett.} {\bf 113}
  (2014) 241302, [\href{http://xxx.lanl.gov/abs/1409.6217}{{\tt
  arXiv:1409.6217}}].

\end{thebibliography}

\providecommand{\href}[2]{#2}\begingroup\raggedright\endgroup
  
\end{document}